\documentclass{ws-mpla}
\usepackage{color}
\usepackage{xspace}
\usepackage{amsmath}
\usepackage{ifthen} 
\newboolean{uprightparticles}
\setboolean{uprightparticles}{false} 

\def\ux85 {\mbox{UX85}\xspace}

\def\babar  {\mbox{BaBar}\xspace}

\ifthenelse{\boolean{uprightparticles}}%
{

 \def\Pmu         {\ensuremath{\upmu}\xspace}

 \def\PDelta      {\ensuremath{\Delta}\xspace}                 
 \def\PXi      {\ensuremath{\Xi}\xspace}                 
 \def\PLambda      {\ensuremath{\Lambda}\xspace}                 
 \def\PSigma      {\ensuremath{\Sigma}\xspace}                 
 \def\POmega      {\ensuremath{\Omega}\xspace}                 
 \def\PUpsilon      {\ensuremath{\Upsilon}\xspace}                 
 
 \def\PB      {\ensuremath{\mathrm{B}}\xspace}                 
                  
 \def\PD      {\ensuremath{\mathrm{D}}\xspace}

 \def\PK      {\ensuremath{\mathrm{K}}\xspace}

 \def\Pb      {\ensuremath{\mathrm{b}}\xspace}

 \def\Pe      {\ensuremath{\mathrm{e}}\xspace}

 \def\Pi      {\ensuremath{\mathrm{i}}\xspace}

 \def\Ps      {\ensuremath{\mathrm{s}}\xspace}

}
{

 \def\Pmu         {\ensuremath{\mu}\xspace}

 \mathchardef\PDelta="7101
 \mathchardef\PXi="7104
 \mathchardef\PLambda="7103
 \mathchardef\PSigma="7106
 \mathchardef\POmega="710A
 \mathchardef\PUpsilon="7107
                  
 \def\PB      {\ensuremath{B}\xspace}                 
                  
 \def\PD      {\ensuremath{D}\xspace}

 \def\PK      {\ensuremath{K}\xspace}

 \def\Pb      {\ensuremath{b}\xspace}

 \def\Pe      {\ensuremath{e}\xspace}

 \def\Pi      {\ensuremath{i}\xspace}

 \def\Ps      {\ensuremath{s}\xspace}

}

\def\epem       {\ensuremath{\Pe^+\Pe^-}\xspace}

\def\mup        {\ensuremath{\Pmu^+}\xspace}
\def\mun        {\ensuremath{\Pmu^-}\xspace} 

\def\squark    {\ensuremath{\Ps}\xspace}

\def\bquark    {\ensuremath{\Pb}\xspace}

\def\kaon  {\ensuremath{\PK}\xspace}
  \def\Kbar  {\kern 0.2em\overline{\kern -0.2em \PK}{}\xspace}

\def\Kz    {\ensuremath{\kaon^0}\xspace}
\def\Kzb   {\ensuremath{\Kbar^0}\xspace}
\def\KzKzb {\ensuremath{\Kz \kern -0.16em \Kzb}\xspace}
\def\Kp    {\ensuremath{\kaon^+}\xspace}
\def\Km    {\ensuremath{\kaon^-}\xspace}

\def\KpKm  {\ensuremath{\Kp \kern -0.16em \Km}\xspace}

  \def\Dbar    {\kern 0.2em\overline{\kern -0.2em \PD}{}\xspace}
\def\D       {\ensuremath{\PD}\xspace}

\def\Dz      {\ensuremath{\D^0}\xspace}
\def\Dzb     {\ensuremath{\Dbar^0}\xspace}
\def\DzDzb   {\ensuremath{\Dz {\kern -0.16em \Dzb}}\xspace}
\def\Dp      {\ensuremath{\D^+}\xspace}
\def\Dm      {\ensuremath{\D^-}\xspace}

\def\DpDm    {\ensuremath{\Dp {\kern -0.16em \Dm}}\xspace}

\def\B       {\ensuremath{\PB}\xspace}
  \def\Bbar    {\kern 0.18em\overline{\kern -0.18em \PB}{}\xspace}

\def\Bu      {\ensuremath{\B^+}\xspace}

\def\Bd      {\ensuremath{\B^0}\xspace}
\def\Bs      {\ensuremath{\B^0_\squark}\xspace}

  \def\Y#1S{\ensuremath{\PUpsilon{(#1S)}}\xspace}

\def\Lbar {\ensuremath{\kern 0.1em\overline{\kern -0.1em\PLambda}}\xspace}

\def\BF         {{\ensuremath{\cal B}\xspace}}

\newcommand{\decay}[2]{\ensuremath{#1\!\to #2}\xspace}         
\def\ra                 {\ensuremath{\rightarrow}\xspace}
\def\to                 {\ensuremath{\rightarrow}\xspace}

\def\bsll     {\decay{\bquark}{\squark \ell^+ \ell^-}}

\def\AT#1     {\ensuremath{A_{\mathrm{T}}^{#1}}\xspace}           

\def\Bsmm     {\decay{\Bs}{\mup\mun}}
\def\Bdmm     {\decay{\Bd}{\mup\mun}}

\def\C#1      {\ensuremath{\mathcal{C}_{#1}}\xspace}                       
\def\Cp#1     {\ensuremath{\mathcal{C}_{#1}^{'}}\xspace}                    
\def\Ceff#1   {\ensuremath{\mathcal{C}_{#1}^{\mathrm{(eff)}}}\xspace}        
\def\Cpeff#1  {\ensuremath{\mathcal{C}_{#1}^{'\mathrm{(eff)}}}\xspace}       
\def\Ope#1    {\ensuremath{\mathcal{O}_{#1}}\xspace}                       
\def\Opep#1   {\ensuremath{\mathcal{O}_{#1}^{'}}\xspace}                    

\newcommand{\tev}{\ensuremath{\mathrm{\,Te\kern -0.1em V}}\xspace}
\newcommand{\gev}{\ensuremath{\mathrm{\,Ge\kern -0.1em V}}\xspace}
\newcommand{\mev}{\ensuremath{\mathrm{\,Me\kern -0.1em V}}\xspace}
\newcommand{\kev}{\ensuremath{\mathrm{\,ke\kern -0.1em V}}\xspace}
\newcommand{\ev}{\ensuremath{\mathrm{\,e\kern -0.1em V}}\xspace}
\newcommand{\gevc}{\ensuremath{{\mathrm{\,Ge\kern -0.1em V\!/}c}}\xspace}
\newcommand{\mevc}{\ensuremath{{\mathrm{\,Me\kern -0.1em V\!/}c}}\xspace}
\newcommand{\gevcc}{\ensuremath{{\mathrm{\,Ge\kern -0.1em V\!/}c^2}}\xspace}
\newcommand{\gevgevcccc}{\ensuremath{{\mathrm{\,Ge\kern -0.1em V^2\!/}c^4}}\xspace}
\newcommand{\mevcc}{\ensuremath{{\mathrm{\,Me\kern -0.1em V\!/}c^2}}\xspace}

\def\mub{\ensuremath{\rm \,\Pmu b}\xspace}

\def\nb {\ensuremath{\rm \,nb}\xspace}

\def\invpb {\ensuremath{\mbox{\,pb}^{-1}}\xspace}

\def\invfb   {\ensuremath{\mbox{\,fb}^{-1}}\xspace}

\def\invps{\ensuremath{{\rm \,ps^{-1}}}\xspace}

\def\gsim{{~\raise.15em\hbox{$>$}\kern-.85em
          \lower.35em\hbox{$\sim$}~}\xspace}
\def\lsim{{~\raise.15em\hbox{$<$}\kern-.85em
          \lower.35em\hbox{$\sim$}~}\xspace}

\def\pt         {\mbox{$p_{\rm T}$}\xspace}

\def\tell1  {TELL1\xspace}
\def\ukl1   {UKL1\xspace}

\newcommand{\eg}{\mbox{\itshape e.g.}\xspace}

\newcommand{\CL}{C.L.\ }
\newcommand{\CLsb}{\ensuremath{\textrm{CL}_{\textrm{s+b}}}\xspace}
\newcommand{\CLs}{\ensuremath{\textrm{CL}_{\textrm{s}}}\xspace}
\newcommand{\CLb}{\ensuremath{\textrm{CL}_{\textrm{b}}}\xspace}

\newcommand{\bb}{\ensuremath{b\bar{b}}\xspace}

\newcommand{\Bsmumu}{\ensuremath{\Bs\to\mu^+\mu^-}\xspace}

\newcommand{\DKpi}{\ensuremath{\D^0\to K^-\pi^+}\xspace}
\newcommand{\Bsmumugamma}{\ensuremath{\Bs\to\mu^+\mu^-\gamma}\xspace}
\newcommand{\Bdpipi}{\ensuremath{\Bd\to\pi^+\pi^-}\xspace}

\newcommand{\BsKK}{\ensuremath{\Bs\to K^+K^-}\xspace}

\newcommand{\BdKpi}{\ensuremath{\Bd\to K^+\pi^-}\xspace}

\newcommand{\Bhh}{\ensuremath{B^0_{(s)}\to h^+h'^-}\xspace}
\newcommand{\Bmm}{\ensuremath{B^0_{(s)}\to \mu^+\mu^-}\xspace}

\newcommand{\BuJpsiK}{\ensuremath{B^-\to J/\psi K^-}\xspace}
\newcommand{\BuJpsimumuK}{\ensuremath{B^-\to J/\psi(\mu^+\mu^-)K^-}\xspace}

\newcommand{\BdJpsiKst}{\ensuremath{B^0_d\to J/\psi K^{*0}}\xspace}

\newcommand{\BcJpsimumumunu}{\ensuremath{B^+_c\to J/\psi(\mu^+\mu^-)\mu^+\nu_{\mu}}\xspace}

\newcommand{\Bqmumu}{\ensuremath{\ensuremath{B^0_{s,d}}\to\mu^+\mu^-}\xspace}
\newcommand{\Bsdmm}{\ensuremath{\ensuremath{B^0_{s,d}}\to\mu^+\mu^-}\xspace}
\newcommand{\BsJpsimumuPhiKK}{\ensuremath{B^0_s\to J/\psi(\mu^+\mu^-) \phi(K^+K^-)}\xspace}

\newcommand{\BsJpsiPhi}{\ensuremath{B^0_s\to J/\psi \phi}\xspace}

\newcommand{\BRof}[1]{\ensuremath{{\cal B}(#1)}\xspace}

\newcommand\TTstrut{\rule{0pt}{3.2ex}}

\newcommand\BBstrut{\rule[-1.8ex]{0pt}{0pt}}

\newcommand{\figref}[1]{Fig.~\ref{#1}}

\begin{document}

\markboth{J. Albrecht}
{Searches for the rare decays \Bsdmm}

\catchline{}{}{}{}{}


\title{Brief review of the searches for the rare decays \linebreak \Bsmm and \Bdmm}

\author{\footnotesize Johannes Albrecht}

\address{CERN\\ 1211 Geneva, Switzerland\\
Johannes.Albrecht@cern.ch}

\maketitle

\pub{Received 18 July 2012\\Accepted 18 July 2012}{Published 13 August 2012}
\begin{abstract}
The current experimental status of the searches for the very rare
decays \Bsmm and \Bdmm is discussed. These channels are highly
sensitive to various extensions of the Standard Model, specially in
the scalar and pseudoscalar sector.  
The recent, most sensitive measurements from the CDF, ATLAS, CMS and LHCb
collaborations are discussed and the combined upper exclusion limit on
the branching fractions determined by 
the LHC experiments is shown to be $4.2\times 10^{-9}$ for \Bsmm and
$0.8\times 10^{-9}$ for \Bdmm.
The implications of these tight bounds on a selected set of New
Physics models is sketched.

\end{abstract}

\ccode{PACS Nos.: 13.20.He 13.30.Ce 12.15.Mm 12.60.Jv}

\newpage
\section{Introduction}

Processes, which are highly suppressed in the Standard Model (SM),
such as decays mediated by flavour changing neutral currents (FCNC) allow
stringent tests of our current understanding of particle physics. 
These transitions are forbidden at tree level in the SM, as all
electrically neutral particles ($\gamma$, $Z^0$, $H^0$ and gluons)
have only diagonal couplings in the flavor space. FCNC processes are
therefore only allowed through loop contributions and probe the
underlying fundamental theory at the quantum level, where they are
sensitive to masses much higher than that of the $b$-quark. 
Historically, many observations have first been indicated by FCNC
processes, examples include the existence of the charm quark or the high
top quark mass.

Precise measurements of the branching fractions of the two FCNC decays
\Bsmm and \Bdmm belong to the most promising modes for a possible
discovery of a theory beyond the SM. 
These decays are strongly suppressed by loop and helicity factors, making the
SM branching fraction small: $3.1\times10^{-9}$ for \Bs decays and
$1\times10^{-10}$ for \Bd decays, both known with a precision better
than 10\%.

%

Enhancements of the decay rates of these decays are predicted in a
variety of different New Physics models. It has been emphasized 
many times that this decay is very sensitive to the presence of
supersymmetric particles\cite{Choudhury:1998ze,Ellis:2005sc,Carena:2006ai,Ellis:2007ss,Mahmoudi:2007gd,Golowich:2011cx,Akeroyd:2011kd,Huang:1998vb,Babu:1999hn}.  
%
For example, in the minimal supersymmetric Standard Model (MSSM), the
enhancement is proportional\cite{Babu:1999hn,Hamzaoui:1998nu,Huang:2000sm} to
$\tan^6\beta$, where $\tan\beta$ is the ratio of the
vacuum expectation values of the two Higgs fields.
For large values of $\tan \beta$, this search belongs to the most
sensitive probes for physics beyond the SM which can be performed at
collider experiments.  
Other models such as non minimal flavor violating or Littlest Higgs
models as well as those with extra dimensions like Randall Sundrum
models predict large effects independent of the value of $\tan
\beta$\cite{Blanke:2006eb,Blanke:2008yr,Liu:2009hv,Bauer:2009cf,Buras:2010cp,Buras:2012ts,Chankowski:2002wr,Arnowitt2002121}.  
In the absence of an observation, limits on \BRof{\Bsmm} are
complementary to those provided by high \pt experiments.
The interplay between both allows to optimally constrain the SUSY
parameter space.


Measuring the decay rates of these decays has been a major goal of
particle physics experiments in the past decade. The limit on the
decay rates was gradually improved by the CDF and D0 experiments at
the Tevatron and the CMS, ATLAS and LHCb experiments at the LHC.   

In this review, the prediction of the branching fraction of
\Bqmumu is discussed in Sec.~\ref{sec:th}. 
The most sensitive measurements of the branching fractions of
\Bsmm and \Bdmm are discussed in Sec.~\ref{sec:exp}, including 
a combination of the measurements performed by the three LHC experiments.
The review is closed by a brief discussion of the implications of these
measurements on various extensions of the Standard Model of particle
physics in Sec.~\ref{sec:impl}.

\section{Theory expectation of the branching fractions}
\label{sec:th}

In this section, the calculation of the branching fraction of \Bsmm
and \Bdmm is first discussed in a model independent way, followed by
the numerical SM prediction.
For simplicity of notation, the expressions are given solely
for the decay \Bsmm. The corresponding expression for \Bdmm decays can
be trivially obtained by exchanging the $s$ with a $d$ quark. 


\subsection{Model independent discussion}

\begin{figure}[t]
\centering
\includegraphics[width=0.49\textwidth]{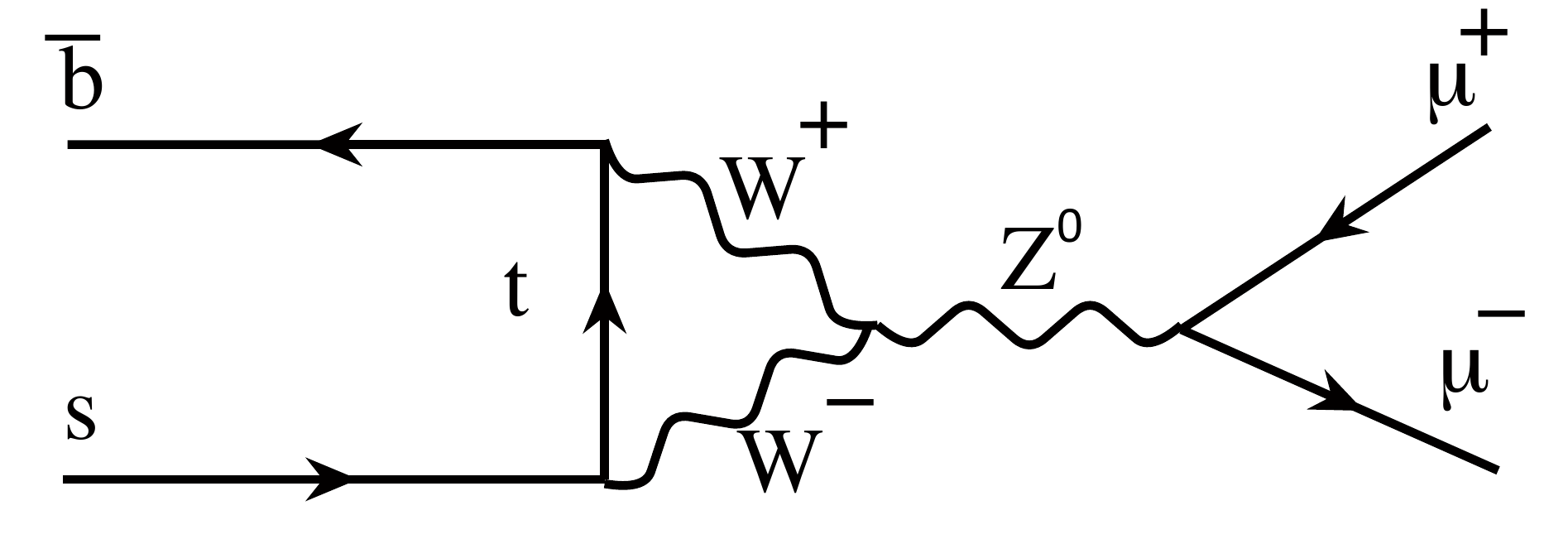}
\includegraphics[width=0.49\textwidth]{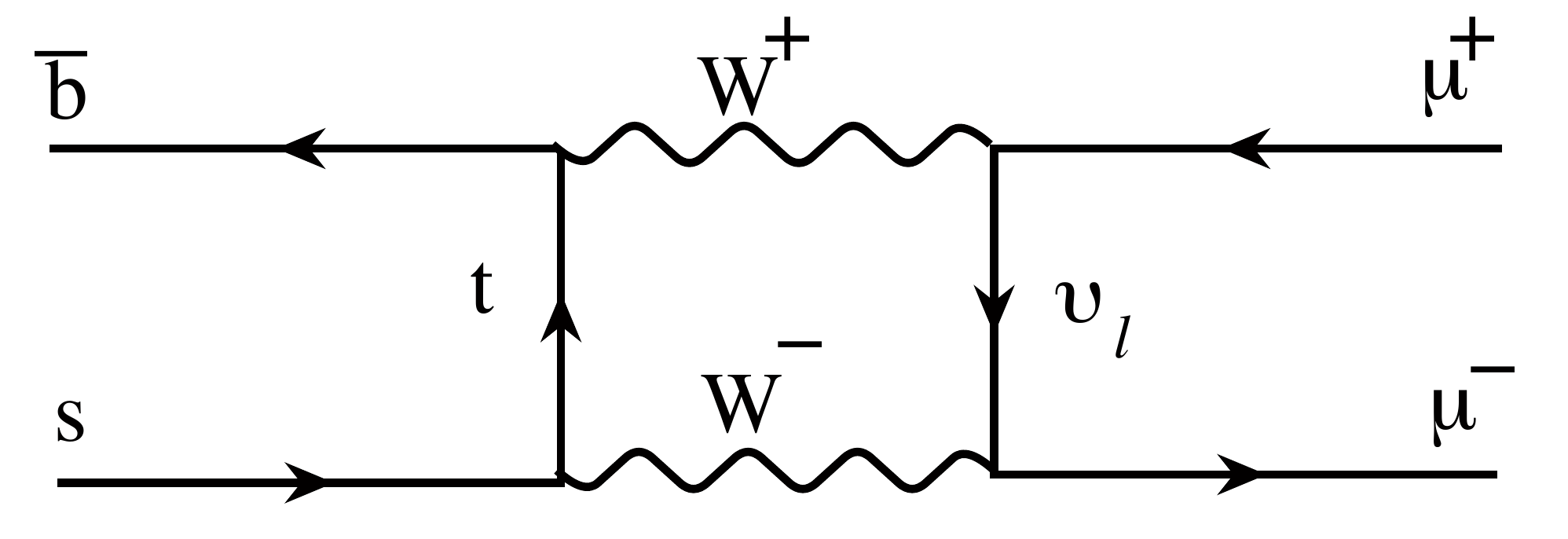}
\caption
{The two dominating SM Feynman diagrams contributing to the decay \Bsmm.}
\label{fig:feynman}
\end{figure}

The branching fraction of \Bsmm can be expressed as low-energy
effective Hamiltonian using the operator product expansion (OPE) which
allows to separate the long distance contributions to the decay
amplitude from the short distance contributions. The former are
relegated to non-perturbative hadronic matrix elements whereas the
latter are described by perturbatively calculable Wilson coefficients
$C_k$. A detailed discussion of the concept can be found in 
Ref.~\refcite{Fleischer:own}.
The effective Hamiltonian for \bsll transitions is given by
\begin{equation}
{\cal{H}}_{eff}=-\frac{4G_F}{\sqrt{2}} V_{tb} V_{ts}^*
\frac{e^2}{16\pi^2}
\sum_i (C_i O_i + C_{i}'O_{i}' ) + h.c. \, ,
\end{equation}
where $G_F$ is the Fermi constant, $V_{tb}$ and $V_{ts}$ are elements
of the Cabibbo-Kobayashi-Maskawa (CKM) matrix and $C_i$ are the Wilson
coefficients. The four-fermion operators $O_i$ which are relevant for
\bsll decays are: 
\begin{eqnarray}
O_{10} &=& (\bar{s} \gamma_\mu P_L b)(\bar{l}\gamma^\mu \gamma_5 l)\, ,\\
O_S &=& m_b(\bar{s}P_Rb)(\bar{l}l)\, ,\\
O_P &=& m_b(\bar{s}P_Rb)(\bar{l}\gamma_5l)\, ,
\end{eqnarray}
where $P_{L,R} = \frac{1}{2}(1\mp \gamma_5)$. The corresponding
operators $O_i'$ are obtained from the $O_i$ operators by replacing
$P_L$ with $P_R$. The notation used here follows the one of
Ref.~\refcite{Bobeth:2001sq} and~\refcite{Altmannshofer:2011gn}.
The branching ratio of \Bsmm decays can then be written in a model
independent way as
\begin{eqnarray}
\BRof{\Bsmm} = \frac{4G^2_F \alpha^2}{64\pi^2} f^2_{B_s}m^3_{B_s} \tau_{B_s}
|V_{tb} V_{ts}^*| 
\sqrt{1-\frac{4m^2_\mu}{m^2_{B_s}}}\label{eq:wilson}\\
\times
\bigg\lbrace 
(1-\frac{4m^2_\mu}{m^2_{B_s}}\,
\big|C_{S}-C'_{S}\big| ^2
+\big|
(C_{P}-C'_{P})
+2\frac{m_\mu}{m_{B_s}}(C_{10}-C'_{10})
\big|^2
\bigg\rbrace\, ,
\nonumber
\end{eqnarray}
where $f_{B_s}$ is the \Bs decay constant, $m_{B_s}$ and $\tau_{B_s}$
are the \Bs mass and lifetime respectively.
The contributions of the scalar ($C_S,~C_S'$) and pseudoscalar
($C_P,~C_P'$) operators enter in the branching fraction without
suppression. As the corresponding Wilson coefficients are still largely
unconstrained, there is significant room for contributions from New
Physics models. 
In the SM, however, they are predicted to be strictly vanishing. The
contribution from electroweak penguin diagrams, contained in the
coefficients $C_{10}$ and $C_{10}'$, is suppressed by a helicity
factor $(m_\mu /m_{\Bs})^2 \approx 4\times 10^{-4}$. 
In the Standard Model, only $C_{10}$ is non-zero and its value is
given by the real coefficient $C_{10}^{SM}$.
It is dominated by a $Z^0$ penguin loop (75\%) and a box diagram
(24\%)\cite{Mahmoudi:2012un}. The two dominant SM Feynman diagrams
contributing to the \Bqmumu decay are given in
Fig.~\ref{fig:feynman}. The sensitivity of this decay to scalar and
pseudo-scalar interactions enables stringent tests of New Physics
models with an extended (pseudo-) scalar sector.  

\subsection{Standard Model prediction}

The numerical value of the Standard Model prediction of the branching
fraction of \Bsmm, as given in Eq.~(\ref{eq:wilson}), can be
calculated as 
\begin{equation}
\BRof{\Bsmm} = 3.1 \pm 0.2 \times 10^{-9}\, , 
\label{eq:bsmm_lattice}
\end{equation}
where the input parameters as discussed in Ref.~\refcite{Buras:2012ts} are used. 
The dominant uncertainty comes from the lattice based calculation of
the \Bs decay constant, which is used here from a recent lattice
average\cite{Laiho:2009eu} as $f_{B_s}=227.7\pm6.2\mev$. The lattice
calculations made significant progress recently and thus made this
very high precision
possible\cite{Gamiz:2009ku,Na:2012kp,Davies:2012qf}. Depending on the
choice of the numerical values of the input parameters for both the
lattice and the experimental quantities, different authors calculate
slightly different branching
fractions\cite{Altmannshofer:2011gn,Mahmoudi:2012un,Charles:2011va,utfit},
up to $3.6\times 10^{-9}$. 
A new lattice average by the FLAG-2 group, which represents all big
lattice collaborations, is expected by the end of 2012, which will
hopefully help to settle these discrepancies in the predicted 
numerical value of
\BRof\Bsmm.

Alternatively to the evaluation of Eq.~(\ref{eq:wilson}), the CKM
dependencies and hadronic uncertainties can be eliminated by
expressing the branching fraction as a function of the 
$\Bs-\bar{\Bs}$ meson mass difference $\Delta M_s$. 
The dependence on the decay constant $f_{B_{s}}$ is exchanged with the
bag parameter $\hat{B}_{B_s}$, which can be calculated with smaller
uncertainties.  
The branching fraction of \Bsmm decays is then
given as\cite{Gamiz:2009ku,Buras:2003td}
\begin{eqnarray}
\BRof{\Bsmm} &=& 4.36\times10^{-10}\times
\frac{\tau_{\Bs}}{\hat{B}_{B_s}}\frac{Y^2(v)}{S(v)}\Delta M_s
\\ 
&=& 3.2 \pm 0.2 \times 10^{-9}\, ,
\label{eq:bsmm_dms}
\end{eqnarray}
where $Y^2(v)$ and $S(v)$ are the two master functions of CMFV
models\cite{Buras:2003jf} , evaluated in the SM with $v=m^2_t/M^2_W$. In
Eq.~(\ref{eq:bsmm_dms}), the numerical value of
$\hat{B}_{B_s}=1.33\pm0.06$ from Ref.~\refcite{Laiho:2009eu}
and~\refcite{Shigemitsu:2009jy} is used. 
It is a remarkable success of the lattice calculations that both
values determined in Eq.~(\ref{eq:bsmm_lattice}) and
Eq.~(\ref{eq:bsmm_dms}) are in excellent agreement.

\subsection{Comparison of the calculated BR with experiment}
\label{sec:thexp}

The LHCb collaboration recently measured the width difference in
the \Bs system in a time dependent analysis of \BsJpsiPhi decays\cite{LHCb-CONF-2012-002} as
$\Delta \Gamma_s =0.116\pm0.019\invps$.
If the calculated value of $\BRof{\Bsmm}$ is compared to a
measurement, the finite width difference has to be taken into account
as the measured branching fraction is the time integrated one, 
\begin{equation}
\BRof{\Bsmm}^{\textrm{exp, }\langle t \rangle} =\frac{1}{2}\int_{0}^{\infty} 
\langle \Gamma(\Bs
(t) \ra \mu^+\mu^-) \rangle dt\, ,
\end{equation}
whereas the the CP averaged branching fraction is calculated. To
compare the measured value with a theory prediction, 
one of the two values has to be corrected, as pointed out
recently\cite{deBruyn:2012wk,deBruyn:2012wj}. Here, a correction of the 
theoretical value (Eq.~(\ref{eq:bsmm_lattice})) is chosen which reads
\begin{equation}
\BRof{\Bsmm}^{\textrm{TH, }\langle t \rangle} = \frac{1}{1-y_s} \times \BRof{\Bsmm}^{CP}
= 3.4 \pm 0.2 \times 10^{-9}\, ,
\label{eq:bsmm_corr}
\end{equation}
where the CP averaged branching ratio as given in
Eq.~(\ref{eq:bsmm_lattice}) is taken and $y_s$ is given as 
\begin{equation}
y_s=\frac{\Delta \Gamma_s}{2\Gamma_s} =0.088\pm0.014\, .
\end{equation}
To compare the SM prediction to a measured branching fraction or an
exclusion limit, the value of the branching fraction of \Bsmm given in
Eq.~(\ref{eq:bsmm_corr}) has to be used.


\section{Experimental situation}
\label{sec:exp}

The search for \Bqmumu has been performed both at the B-factories
\babar and Belle as well as at the experiments at the Tevatron and at
the LHC. At the B-factories, running at the
$\Upsilon(4S)$ resonance, only \Bd decays are accessible whereas at
the Tevatron and the LHC both \Bd and \Bs mesons are produced. 

A comparison of the the \bb-production cross section at the different
experiments is given in Tab.~\ref{tab:xsec}. The efficiency of the
acceptance and \pt requirements is not corrected for, as the applied
selections are typical for analyses at the given experiments.
The production cross section at hadron colliders is, depending on
collision energy and detector acceptance, between 6\mub (CDF, D0) and
94\mub (LHCb).  Note that the LHCb experiment is a single arm, forward
spectrometer which integrates the peak of \bb-pairs produced in
forward direction whereas the general purpose detectors ATLAS and CMS
(GPD) have a limited acceptance in the forward or backward region but
integrate over the 
complete remaining phase space. The effect of these acceptances is
that the \bb-production cross section is similar for
the LHCb experiment and the GPD. The sensitivity of recent HEP
experiments for the searches for \Bqmumu can be summarized as follows: 
\begin{table}[b]
\tbl{\bb-production cross sections in the acceptance of recent HEP
  experiments. Data taken from
  Refs.~\protect\refcite{babarPhys,PhysRevD.75.012010,Aaij:2012jd,Chatrchyan:2011pw}. 
  The cross section for ATLAS, CMS, CDF, LHCb is calculated from a $B^+$
  cross section measurement with \pt$>5\gev$ (CDF: \pt$>6\gev$, LHCb
  \pt$>0\gev$) using the LHCb measurement of the hadronization
  fractions\protect\cite{PhysRevD.85.032008}. The number of produced \bb-pairs
  estimated for the dataset recorded until the end of 2011 and does
  not include trigger and reconstruction efficiencies.}   
{\begin{tabular}{cccccc}
\hline
\TTstrut\BBstrut
~experiment~&~acceptance~&~$\Upsilon
(4S)$~&~$p\bar{p},~1.96\tev$~&~$pp,~7\tev$~& \bb-pairs produced\\
\hline    
\TTstrut
\babar, Belle&$4\pi$    &1.05\nb&&&$\sim 1\times 10^9$\\
CDF&$| \eta  |<1 $&&$6.3\pm0.6\mub$&&$\sim 6 \times 10^{10}$\\
ATLAS, CMS&$| \eta |<2.2 $&&&$75\pm17\mub$&$\sim4\times 10^{11}$\\
\BBstrut LHCb&$2<\eta<6$&&&$94\pm8\mub $&$\sim 9\times 10^{10}$\\
\hline
 \end{tabular}
\label{tab:xsec}}
\end{table}
%
%
\begin{itemize}
\item Compared to the cross sections at the Tevatron and the LHC, the
  \bb-production rate at \epem colliders at $\Upsilon(4S)$ is more
  than three orders of magnitude smaller. Hence, the number of
  recorded \bb-pairs is much lower and the searches for \Bqmumu are
  with the existing datasets of 500 and 1000\invfb not
  competitive. Therefore, the discussion in this review focuses on the  
  hadron collider experiments.

\item The Tevatron collider experiments recorded a dataset
  corresponding to an integrated luminosity of about 10\invfb in Run~II between
  2001 and 2011, with an estimated number of $6\times10^{10}$
  \bb-pairs produced per experiment.

\item
Since the LHC began collecting data in 2010, its experiments have
taken over the lead of the field due to the very high \bb-production
cross section.
At the general purpose detectors at the LHC, about $4\times10^{11}$
\bb-pairs have been produced in their dataset of 5\invfb which has
been collected at instantaneous luminosities up to $3.5\times
10^{33}$\,s$^{-1}$m$^{-2}$. The LHCb experiment, in contrast to the
GPD, leveled its luminosity in 2011 to a constant value of $2-3\times
10^{32}$\,s$^{-1}$m$^{-2}$, 
giving a total dataset until the end of 2011 of 1\invfb which contains
$9\times 10^{10}$ \bb-pairs.
\end{itemize}
Thanks to the very large number of \bb-pairs produced and the clean
experimental signature of the \Bqmumu decays, the hadron
collider experiments have performed the most sensitive searches for
the decays \Bqmumu. These are discussed in detail in the remainder of
this section.

\subsection{Overview of the analyses and search strategies}

The search for the rare decays \Bqmumu has been performed at both
Tevatron detectors.
The analyses discussed here contain the published CDF analysis using
7\invfb of $p\bar{p}$ collision data\cite{Aaltonen:2011fi} at $\sqrt s =
1.96\tev$ as well as their unpublished update\cite{cdf_10fb} using the
full dataset of 10\invfb, collected until the shutdown of the Tevatron
collider. The D0 collaboration published a result\cite{Abazov:2010fs}
with a dataset corresponding to an integrated luminosity of
6.1\invfb. 
The sensitivity of the D0 detector is about a factor of two
worse than the CDF sensitivity as the D0 mass resolution is much worse. For
the decay \Bsmumu, a mass resolution of 24\mevcc is expected for the
CDF experiment whereas 120\mevcc is expected for the D0 experiment.
Due to its lower sensitivity, the D0 result is not discussed further
in this review.  

At the LHC, the most sensitive measurements of the branching ratios of
\Bqmumu have been made by the LHCb collaboration, which published three
measurements\cite{Aaij:2011rja,LHCb:2011ac,Aaij:2012ac} using 37\invpb,
370\invpb and 1\invfb of $\sqrt s =
7\tev$ $pp$-collision data, respectively. 
Both general purpose detectors at the LHC have performed searches for
\Bqmumu, the ATLAS collaboration\cite{Aad:2012pn} using 2.4\invfb of
data and the CMS
collaboration\cite{Chatrchyan:2011kr,Chatrchyan:2012rg}, using 1\invfb
and 5\invfb of $\sqrt s =7\tev$ $pp$-collision data.   

The invariant mass resolution of the three LHC experiments, estimated
for \Bsmm signal candidates, is
\begin{eqnarray}
{\rm ATLAS} &:& 60\mevcc~(|\eta|<1) - 110\mevcc~(|\eta|>1.5)\, ,\nonumber \\
{\rm CMS} &:& 37\mevcc~(|\eta|\sim 0 ) - 77\mevcc~(|\eta|>1.8)\, ,\nonumber\\
{\rm LHCb} &:& 24.8\mevcc\, ,\nonumber
\end{eqnarray}
which shows the invariant mass resolution of the LHCb forward
spectrometer to be on average a factor two better than the CMS
resolution and a factor 3-4 better than the ATLAS resolution. This, together
with the better separation between signal and background in the boosted
forward region and the cleaner environment compensates for the
luminosity, which is a factor 5 lower in LHCb compared to the GPDs.

The measurements performed by the CDF and LHCb collaborations use a
similar strategy to search for \Bqmumu decays. 
Triggered opposite sign dimuon candidates are further selected to
clean up the sample.
To maximize the efficiency, a significant amount of background is
allowed in the selection. The remaining events are classified in
a two-dimensional plane spanned by the invariant mass and the signal
likelihood formed from the two-prong decay signature. 

In contrast to this, the ATLAS and CMS collaborations use a tight
selection which is cut 
based at CMS and based on a boosted decision tree in
ATLAS. Both measurements are binned in rapidity, separating 
regions with different detector performance and therefore
different signal to background ratios.

The event selection and classification with a signal likelihood is
discussed in Sec.~\ref{sec:sel}.
All experiments use a relative normalization to interpret the observed
pattern of events as branching ratio, discussed in Sec.~\ref{sec:norm},
and then use the modified frequentist method (\CLs) to extract the
limits on the branching fractions, as discussed in Sections~\ref{sec:cls}
and \ref{sec:result}. In Sec.~\ref{sec:comb}, a combination of the
searches for \Bqmumu of the three LHC experiments is discussed.

\subsection{Event selection and Signal Likelihood} 
\label{sec:sel}

All experiments have highly efficient muon triggers, which are used
both to select events for the signal and for the normalization
channel (see, \eg, Refs.~\refcite{Aad:2012xs} or~\refcite{Aaij:1384386} for
a detailed discussion). The dimuon triggers for the \Bsmm analysis
obtained a high priority in the physics programs of the GPDs,
therefore, the high efficiency could be preserved during the high
luminosity running. 

The fully reconstructed signal candidates can be identified by their
clearly separated secondary vertex, exploiting the long \Bs meson
lifetime. The candidate momentum vector will be aligned with the
separation between primary and secondary vertex (pointing), and the two
muon candidates will be isolated from other tracks in the event, due
to the hard $B$ fragmentation. 

Background events tend to be partially reconstructed and shorter lived
than the signal. They also have a softer \pt spectrum, lower degree of
isolation and a less precise pointing. The dominant
combinatorial background consists of sequential semileptonic decays
($b\ra c\mu^-\ra \mu^+\mu^- X$) and double semi-leptonic decays ($b\ra\mu^+
X$ , $b\ra\mu^- X$). The former mostly populate the lower mass
sidebands whereas the latter cover the whole mass range.
In the analyses, these combinatorial backgrounds are estimated by an
extrapolation from the invariant mass sidebands.

Peaking backgrounds, dominantly originating from misidentified \BsKK,
\Bdpipi and \BdKpi 
decays, need to be evaluated separately. Here, the invariant mass
line shape can be extracted from simulated samples of doubly
misidentified \Bhh events. The misidentification rates
$K^\pm\ra\mu^\pm$ and $\pi^\pm \ra \mu^\pm$ are extracted in data from
control channels such as \DKpi. The contributions of exclusive decays
such as \BcJpsimumumunu or \Bsmumugamma events is found to be
negligible. 


\subsubsection{Classification of events}

The CDF and LHCb measurements use a loose selection that allows
significant amounts of background and then use a multivariate selection to
discriminate signal from background. In this review only the
more sensitive analysis of the LHCb collaboration is discussed in detail.

\vskip 5mm \noindent
\subsubsection*{LHCb analysis of 1\invfb}

The selected sample is first cleaned with a multivariate
classifier based on six variables that can be used equally for
the signal and the normalization channels. This selection removes 80\% of
the residual background, while retaining 92\% of the signal.
Applying it improves the performance of the main multivariate
selection described below.

About 17\,000 dimuon candidates pass this selection while 11.6 \Bsmm and
1.3 \Bdmm candidates are expected, assuming SM rates.
The selected dimuon candidates are classified in a binned two-dimensional
space formed by the dimuon invariant mass and the output of another
boosted decision tree (BDT) which combines nine variables to optimally
exploit geometrical and kinematic information. 
The variables used to construct the BDT include the $B$ candidate
pointing, decay time and \pt, the displacement of the muons from the
primary vertex and the distance between the muon pairs as well as a
measure of the track and $B$ candidate isolation. 

No data were used for
the choice of the variables and the subsequent training of
the BDT, to avoid biasing the results. Instead the BDT
was trained using simulated samples, \Bsmm for signal and $\bb \ra
\mu^+ \mu^- X$ for background. It is defined such
that for the signal it is approximately uniformly distributed between
zero and one, while for the background it peaks at zero. 
Of the 17\,321 dimuon candidates after the selection\footnote{The
 number of events quoted here are in a mass range
between 4900\mevcc and 6000\mevcc.}, only 95 are in
the region of high signal likelihood (BDT$>0.5$) which contains about
50\% of the expected signal.

The probability for a signal event to have a given BDT value is
obtained from data using an inclusive \Bhh sample. Only events
triggered independently of the signal candidates are considered. The
fraction of \Bhh signal candidates in each bin of BDT is determined by
fitting the $h^+h'^-$ invariant mass distribution.

The chosen number and size of the bins are a compromise between
maximizing the number of bins and the necessity to have enough \Bhh
events to calibrate the signal BDT and enough background in the mass
sidebands to estimate the combinatorial background in the
\Bs and \Bd mass regions. The BDT range is thus divided into eight
bins and the invariant mass range into nine bins.

\vskip 5mm \noindent
\subsubsection*{CDF analysis}

The CDF collaboration uses a similar multivariate classifier (called
$\nu_N$) which is based on an artificial neural net combining 14
input variables. The sample is binned in 8 bins of $\nu_N$ and 5 bins
of invariant mass. The signal shape of $\nu_N$ is taken from the
simulation and cross checked with \BuJpsiK events.

The evaluation of the compatibility of the observed pattern of events
with the background or signal hypotheses is then performed in these
bins and will be described in detail in Sec.~\ref{sec:cls} and~\ref{sec:result}.

\subsubsection{Cut based separation of signal and background}

The ATLAS and CMS measurements use a tight selection of the signal on
which the background is cut away. The ATLAS selection is based on a
boosted decision tree combining 14 variables. To ensure that the data
are reproduced, the simulation is tuned with an iterative reweighting
procedure. 

The CMS collaboration optimizes a cut based selection on MC signal and
data sideband events. 
A multivariate analysis is in preparation which is expected to give an
improvement of up to 20\% relative to the cut based analysis.




\subsection{Relative Normalization}
\label{sec:norm}

The branching fraction for the \Bmm signal is measured by all analyses
discussed here relative to a
channel with a well known branching fraction as
\begin{equation}
{\cal B} =  {\cal B}_{\rm norm} 
\times\frac{\rm \epsilon_{norm} }{\rm\epsilon_{sig}}
\times\frac{f_{\rm norm}}{f_{d(s)}}
\times\frac{N_{\Bmm}}{N_{\rm norm}} \, ,
\label{eq:normalization}
\end{equation}
where the branching fraction is determined by ${\cal B}_{\rm norm}$,
the branching fraction of the normalization channel. It needs to be
corrected by the relative efficiencies $\rm \epsilon_{norm}
/\rm\epsilon_{sig}$, the probabilities that a $b$ quark
fragments into a $B^0_{(s)}$ and into the $b$ hadron involved for the
chosen normalization mode $f_{\rm norm}/f_{d(s)}$ and by the ratio of
observed candidates in the signal and the normalization mode,
$N_{\Bmm}/N_{\rm norm}$. 
This relative normalization has many advantages over a absolute
normalization: the luminosity and production cross section is not
required to be known and the determination of relative efficiencies is
more robust than the determination of absolute efficiencies.

All experiments discussed here measure \BRof{\Bmm} relative to the
\BuJpsimumuK channel, which has a similar muon identification and trigger,
which minimizes systematic uncertainties. Its branching
fraction is known to a precision of 3.5\%\cite{pdg10}.
However, as it is a
three body decay, the detector acceptance as well as the kinematic
distributions of the tracks differ and need to be corrected for. 

The
LHCb collaboration performs the only measurement that uses 
the weighted average of three normalization channels, additionally
\BsJpsimumuPhiKK and \Bhh. 
These three channels have different advantages as normalization: the
\BsJpsiPhi decay is a \Bs decay and hence the fraction of
hadronization probabilities cancels. However, the branching fraction of
this decay is currently known to a precision of only
26\%~\cite{Louvot:2009xg}. The \Bhh channel has an identical two body
signature as the signal decay, but has been selected by very different hadronic trigger
than the dimuon signal. Therefore, only \Bhh candidates that are
unbiased by the trigger selections are used for the normalization. 
While the normalization of the average of three channels significantly
improves the robustness of the analysis and reduces the dependence on
the simulation, the \Bu channel strongly dominates the average.

\begin{figure}[!htb]
\centering
\includegraphics[width=0.47\textwidth,height=0.32\textwidth]{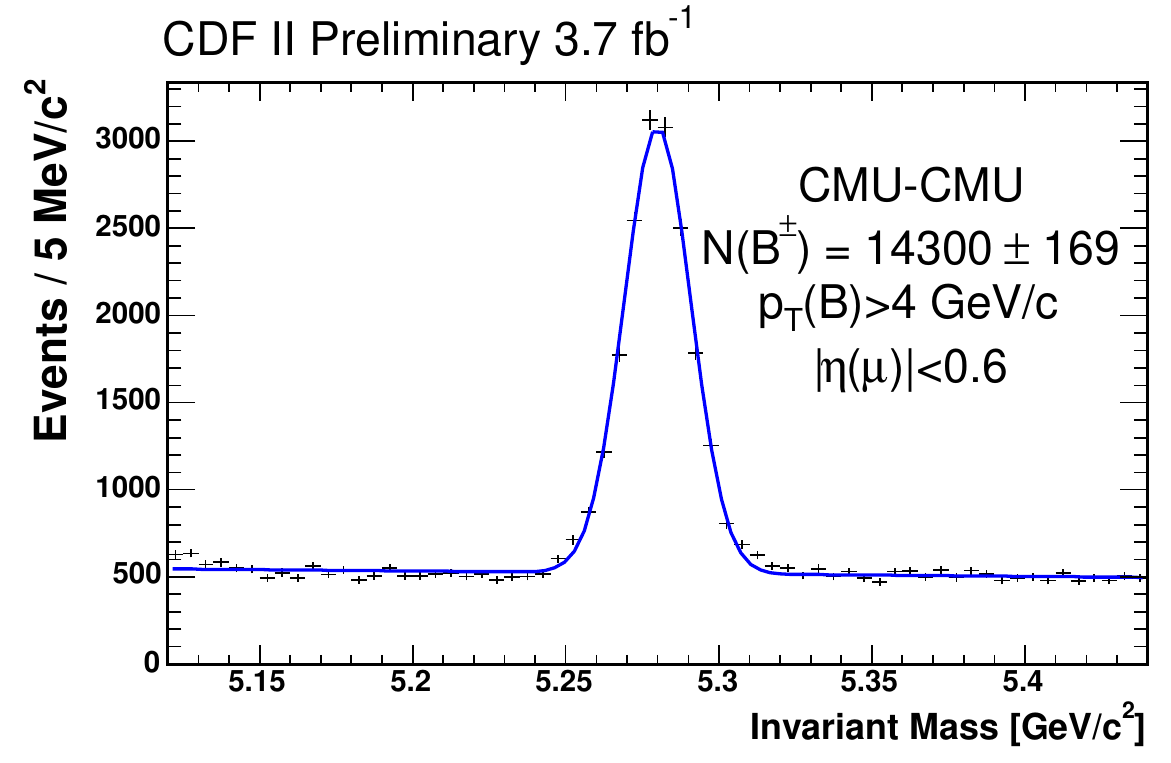}
\includegraphics[width=0.45\textwidth,height=0.31\textwidth]{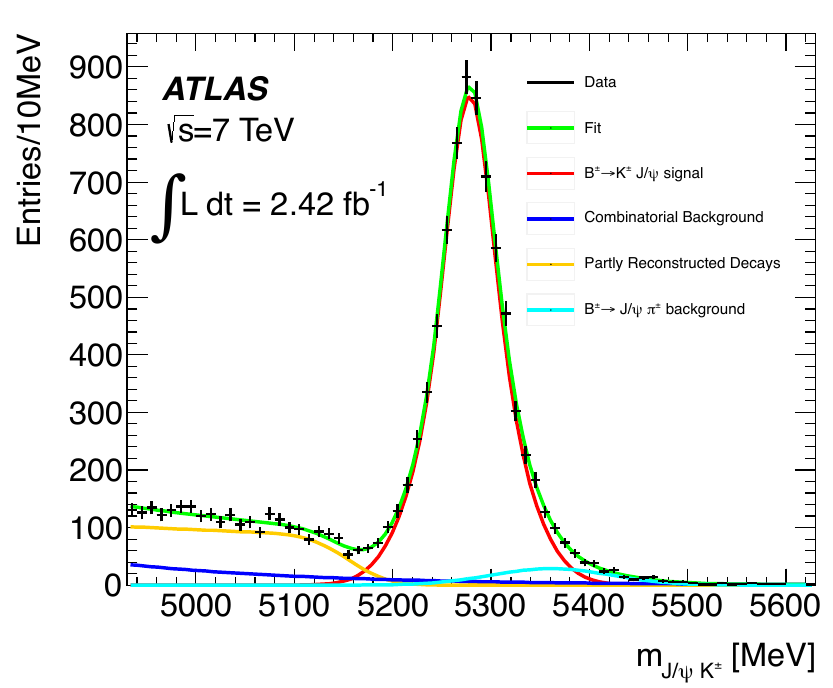}\\
\includegraphics[width=0.47\textwidth,height=0.32\textwidth]{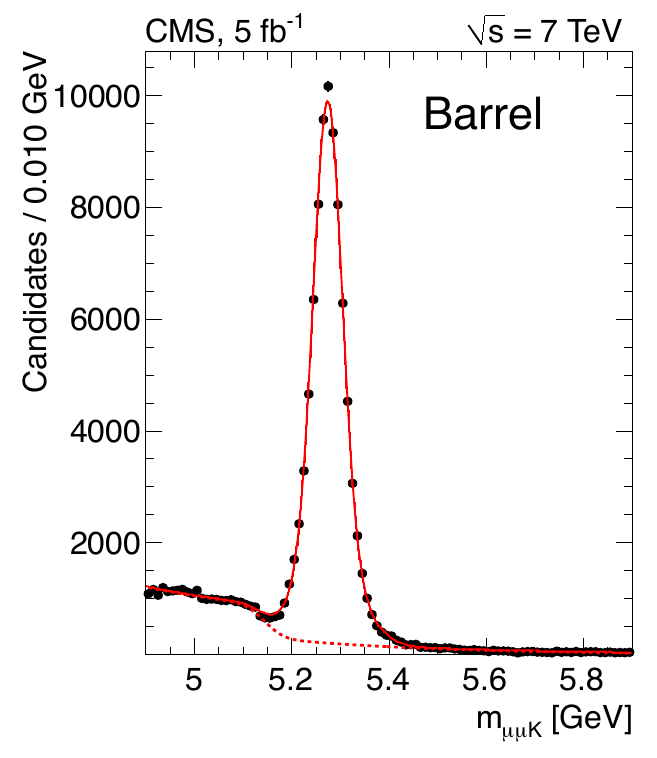}
\includegraphics[width=0.47\textwidth]{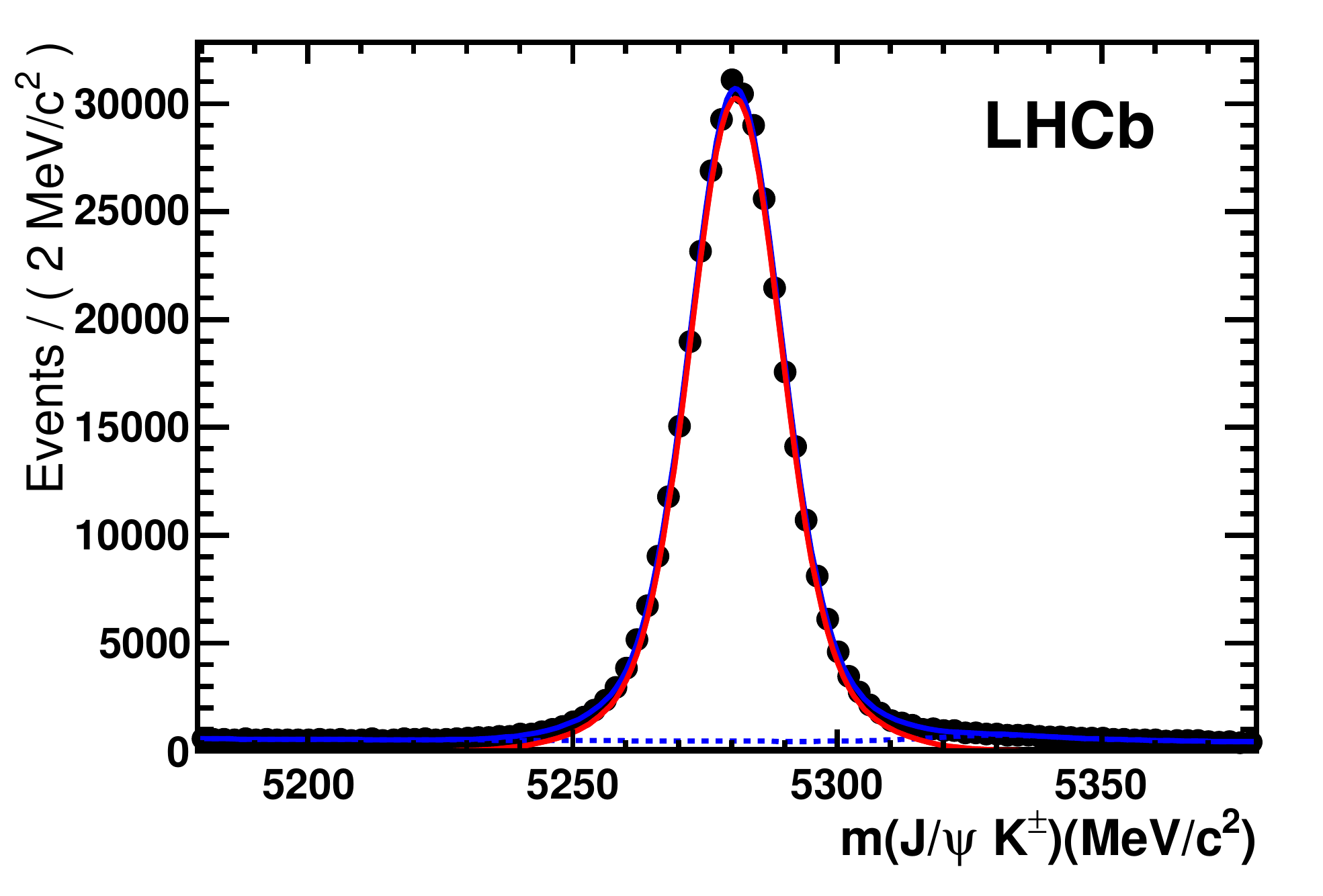}
\caption
{Invariant mass distribution of the \BuJpsiK candidates used in the
  normalization procedure. Note the different scale on the Figures. Reproduced from 
  CDF\protect\cite{cdf_10fb}, ATLAS\protect\cite{Aad:2012pn},
  CMS\protect\cite{Chatrchyan:2012rg} and LHCb\protect\cite{Aaij:2012ac}.}
\label{fig:jpsik}
\end{figure}

The mass distributions of the selected \BuJpsiK candidates are shown in
Fig.~\ref{fig:jpsik} for the four experiment discussed here. 
Please note the different scale of the figures, while the resolution
of the LHCb and CDF experiments is good enough to clearly separate the
shoulder of partially reconstructed \BdJpsiKst events from the signal,
the ATLAS and CMS experiments need to model it.

The relative efficiencies are determined by all experiments from the
simulation and have been verified with \BsJpsiPhi, \BdJpsiKst and
\Bhh events in data. The uncertainties the efficiency ratio
contributes to the normalization factors vary between 3\% (LHCb) and
16\% (ATLAS end-cap).

For the normalization to \Bu candidates, the ratio of
$b$-hadronization fractions need to be included in
Eq.~\ref{eq:normalization}. The CDF experiment uses the PDG
average for the hadronization fraction\cite{pdg10},
which has an uncertainty of 10\%. The LHC experiments all use the
measurement performed by the LHCb collaboration\cite{PhysRevD.85.032008}, which has  
an uncertainty of 7.8\%. The ATLAS and CMS collaborations use this
measurement assuming it is valid for all momentum and rapidity
regions.   




\subsection{Extraction of the branching ratio}
\label{sec:cls}

The branching fraction limit is extracted from the pattern
of events using the \linebreak modified frequentist method\cite{Junk:1999kv,0954-3899-28-10-313}
(\CLs), where the background only (B) and signal plus
background (S+B) hypotheses are tested. The logarithmic form of the
likelihood ratio
\begin{equation}
-2 \ln Q=-2 \ln \frac{{\cal{L}}_{S+B}}{{\cal{L}}_{B}}=
-2 \ln \prod_{i}\frac{e^{s_i+b_i}(s_i+b_i)^{d_i} / d_i!}
{e^{-b_i}(b_i)^{d_i} / d_i!}\, ,
\end{equation}
depends on the expected number of signal ($s_i$) and background
($b_i$) events in each bin $i$, as well as on
the observed number of events ($d_i$). The method provides a measure
of the compatibility of the observed distribution with the signal plus
background hypothesis, \CLsb, as well as a measure for the
compatibility with the background only hypothesis, \CLb, which are both
derived from pseudo experiments. 
The fraction of signal plus background (background only) events 
with a value of $-2\ln Q$ smaller than or equal to the observed
valve is computed. Systematic uncertainties are included in the pseudo experiments
as nuisance parameters using Poisson distributions for the
uncertainties with statistical nature and bifurcated Gaussians for the
others. 

The ratio of
both confidence levels, $\CLs= \CLsb/\CLb$, is used to set an exclusion
limit on the branching fraction of the two decays to protect against
negative statistical fluctuations of the background which could lead
to an exclusion of the null hypothesis without experimental
sensitivity. 
To claim evidence of the decays, $1-\CLb$ is used as a p-value. A 95\%
confidence level exclusion corresponds to a \CLs value of 0.05 and a
3$\sigma$ evidence to a value of $1-\CLb = 2.7\times 10^{-3}$ for a one
sided definition of the significance.

\subsection{Results}
\label{sec:result}
The CDF collaboration observes an excess of candidates in the
\Bsmm signal region. It is concentrated in bins with a high signal
likelihood, $\nu_N > 0.97$. There is also an excess in the medium 
likelihood bins, $0.97 < \nu_N < 0.987$, which appears to be a statistical
fluctuation of the background as there is no significant expectation
of \Bsmm signal consistent with the observation in the two highest 
bins. The p-value for a background-only hypothesis is
0.94\%, when considering only the two highest $\nu_N$ bins the p-value
becomes 2.1\%.  
The branching fraction corresponding to the observed pattern of events
is determined from a log-likelihood fit and is determined to be 
\begin{equation}
\BRof\Bsmm ^{\rm CDF}= 13 ^{+9} _{-7} \times 10^{-9}\, .
\end{equation}
Additionally, bounds are set on the branching fraction using the \CLs
method as \BRof\Bsmm $< 31 \times 10^{-9}$ at 95\% CL. 


The excess seen by the CDF collaboration has not been confirmed by the
LHC experiments, which are summarized in Tab.~\ref{tab:results}. The ATLAS
collaboration analyses 2.4\invfb of data and observes no excess of
candidates and determines an upper limit of \BRof\Bsmm$<22\times
10^{-9}$, being very close to the value expected from background
extrapolation.   

\begin{table}[h]
\tbl{Expected and observed upper limit on \BRof\Bqmumu for the most
  recent measurements, all given at 95\% CL.
  For the expected limits, it is indicated if the SM signal plus
  background or the background only case is assumed.
  The CDF collaboration observes an excess of \Bsmm
  candidates and quotes bounds at 95\% CL on the branching fraction:
  $0.8\times 10^{-9} < \BRof\Bsmm<34\times 10^{-9}$. 
} 
{\begin{tabular}{cccccc}
\hline
\TTstrut
~~~~ experiment~~~~ &~~~~  luminosity~~~~  & \multicolumn{2}{c}{\Bsmm} &\multicolumn{2}{c}{\Bdmm}\\
&& ~~~expected&\textbf{observed}~~~& ~~~expected&\textbf{observed}~~~\\
\BBstrut&& \multicolumn{4}{c}{$\times 10^{-9}$}\\
\hline    
\TTstrut
D0&6\invfb      &40~$^{\rm (bkg)}$&\textbf{51}&-&-\\
\TTstrut CDF&10\invfb    & 13~$^{\rm (bkg)}$ & \textbf{31}  &40~$^{\rm (bkg)}$&\textbf{46}\\
\TTstrut ATLAS&2.4\invfb & 23~$^{\rm (bkg)}$ &\textbf{22} &-&-\\
\TTstrut CMS & 4.9\invfb & 8.4~$^{\rm (SM+bkg)}$&\textbf{7.7}&1.6~$^{\rm (bkg)}$&\textbf{1.8}\\
\TTstrut\BBstrut LHCb & 1\invfb  & 7.2~$^{\rm (SM+bkg)}$&\textbf{4.5}&1.1~$^{\rm (bkg)}$ &\textbf{1.0}\\
\hline
 \end{tabular}
\label{tab:results}}
\end{table}

\begin{figure}[!htb]
\centering
\includegraphics[width=0.8\textwidth,height=0.37\textwidth]{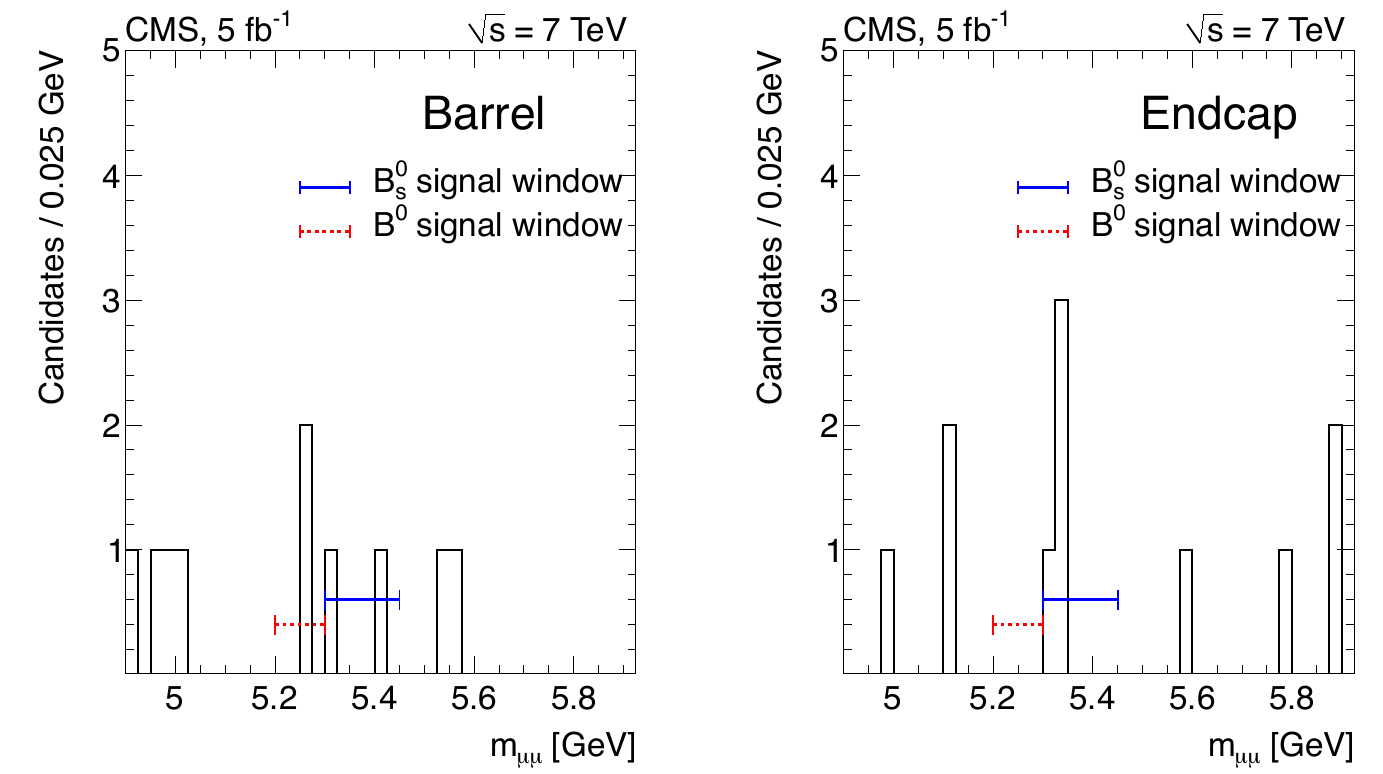}\\
\includegraphics[width=0.8\textwidth]{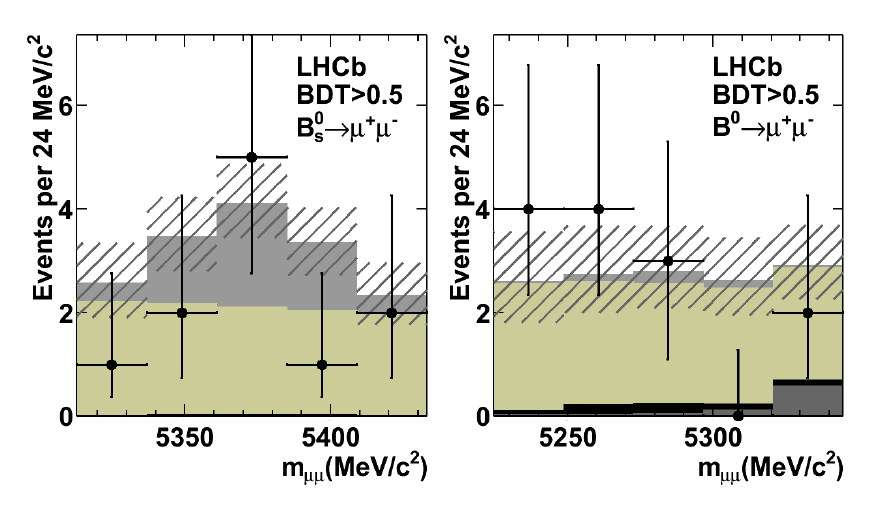}
\vspace{-4mm}
\caption
{ Distribution of selected signal candidates.
  Top: Events observed in CMS in the barrel (left) and end-cap (right)
  channels. 
  Bottom: Events observed in LHCb in the \Bs channel (left) and the
  \Bd channel (right) for BDT$ > 0.5$ and expectation for, from top, SM signal
  (gray), combinatorial background (light gray), \Bhh background
  (black) and cross feed between both modes (dark gray). The hatched
  area depicts the uncertainty on the total background expectation.
  Figures reproduced from CMS\protect\cite{Chatrchyan:2012rg} and
  LHCb\protect\cite{Aaij:2012ac}.}  
\label{fig:signal}
\end{figure}

The CMS collaboration analyses 4.9\invfb of data and observes the
pattern of events shown in Fig.~\ref{fig:signal}~(top). Six events are
observed in the \Bsmm signal region, of which two are in the barrel
and four in the end-cap. This pattern of events is consistent with the
expectation from the Standard Model branching fraction plus
background. The upper limits combined for both regions are at 95\% CL:  
\begin{eqnarray}
\BRof\Bsmm ^{\rm CMS} &<& 7.7\times10^{-9} \textrm{ and }\\
\BRof\Bdmm ^{\rm CMS} &<& 1.8\times10^{-9}\, ,
\end{eqnarray}
which is well comparable to the median expected upper limits of
$8.4\times10^{-9}$ for the \Bs mode and $1.6\times10^{-9}$ for \Bd
mode.

The LHCb collaboration analyses 1\invfb of data and observes the
distribution of the invariant mass for BDT$>0.5$ as shown in 
Fig.~\ref{fig:signal}~(bottom) for \Bs and \Bd decays. 
The expected limits are computed allowing the presence of \Bsmm events
according to the SM rate, including cross feed between the two
modes. A limit of $7.2 (1.1) \times10^{-9}$ is expected for the \Bs
and \Bd modes respectively. The analysis of the observed pattern of
events gives
\begin{eqnarray}
\BRof\Bsmm ^{\rm LHCb} &<& 4.5\times10^{-9} \label{eq:lhcbBs}\textrm{ and }\\
\BRof\Bdmm ^{\rm LHCb} &<& 1.0\times10^{-9}\label{eq:lhcbBd}\, .
\end{eqnarray}
This is a downward fluctuation of $1\sigma$ for the \Bs mode with
respect to the SM expectation.
A simultaneous unbinned likelihood fit to the mass projections in the
eight BDT bins has been performed to 
determine the \Bsmm branching fraction. 
The fit gives 
\begin{equation}
\BRof\Bsmm ^{\rm LHCb}= 0.8 ^{+1.8} _{-1.3} \times 10^{-9}\, ,
\end{equation}
where the central value is extracted from the maximum of the logarithm 
of the profile likelihood and the uncertainty contains both
statistical fluctuations and systematic uncertainties.

In comparison to the CMS measurement, LHCb has a 15\% higher
sensitivity which shows that the better detector resolution (both in
mass and vertex precision) and the more advanced analysis overbalance
the factor five higher luminosity that the CMS experiment has
recorded. Due to a downward fluctuation of the data, the measured
upper exclusion limit on the decay \Bsmm is about 40\% below the CMS
limit. 

\subsection{Combination of  \Bsdmm measurements}
\label{sec:comb}

The results of the analyses performed on the 2011 LHC dataset of the
ATLAS, CMS and LHCb collaborations have been
combined\cite{LHCb-CONF-2012-017}. As the LHCb collaboration has also
published a measurement using the data collected in 2010, the
measurements performed on both datasets are used in the combination.   

\begin{figure}[t]
\centering
\includegraphics[width=0.47\textwidth]{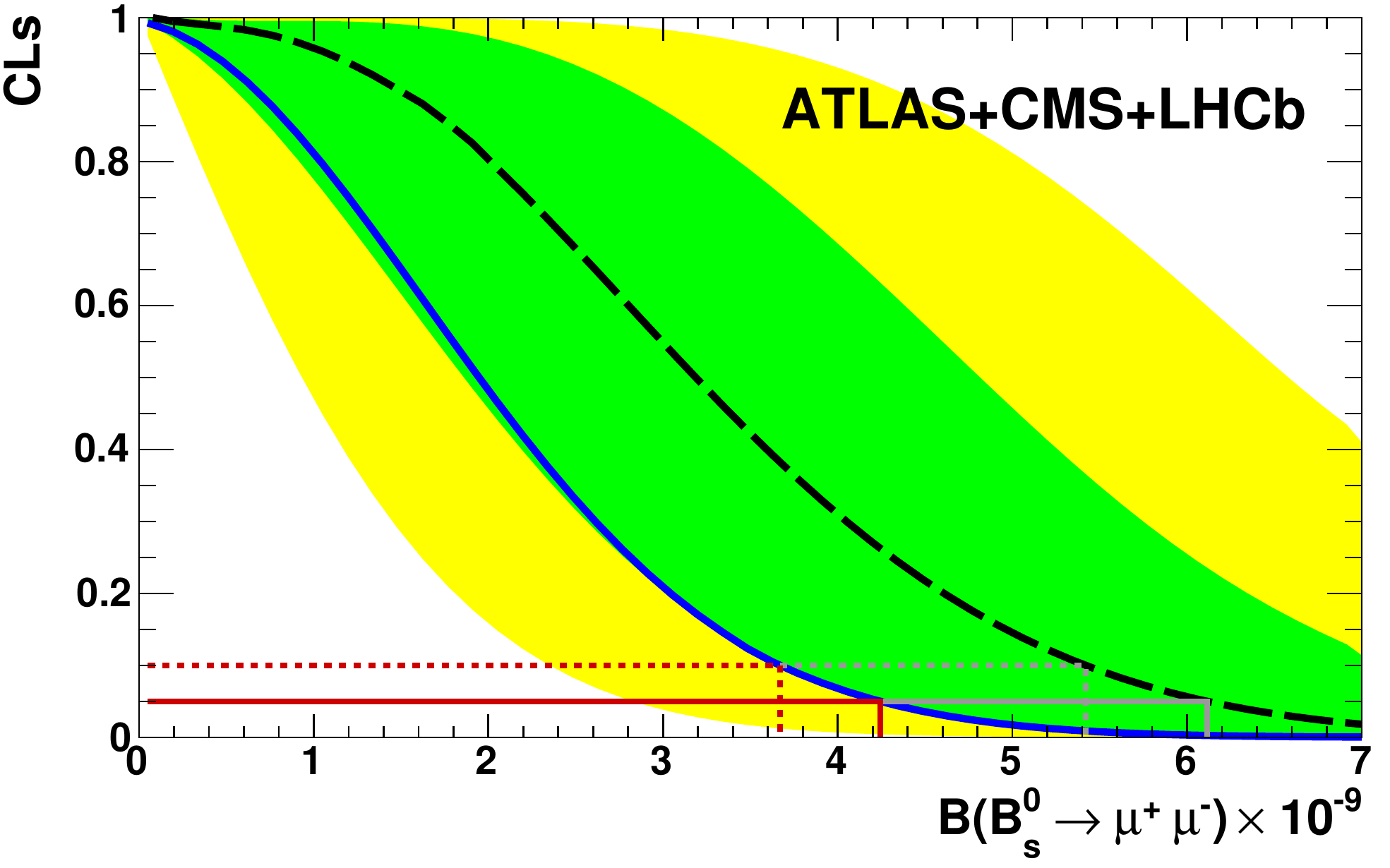}
\includegraphics[width=0.47\textwidth]{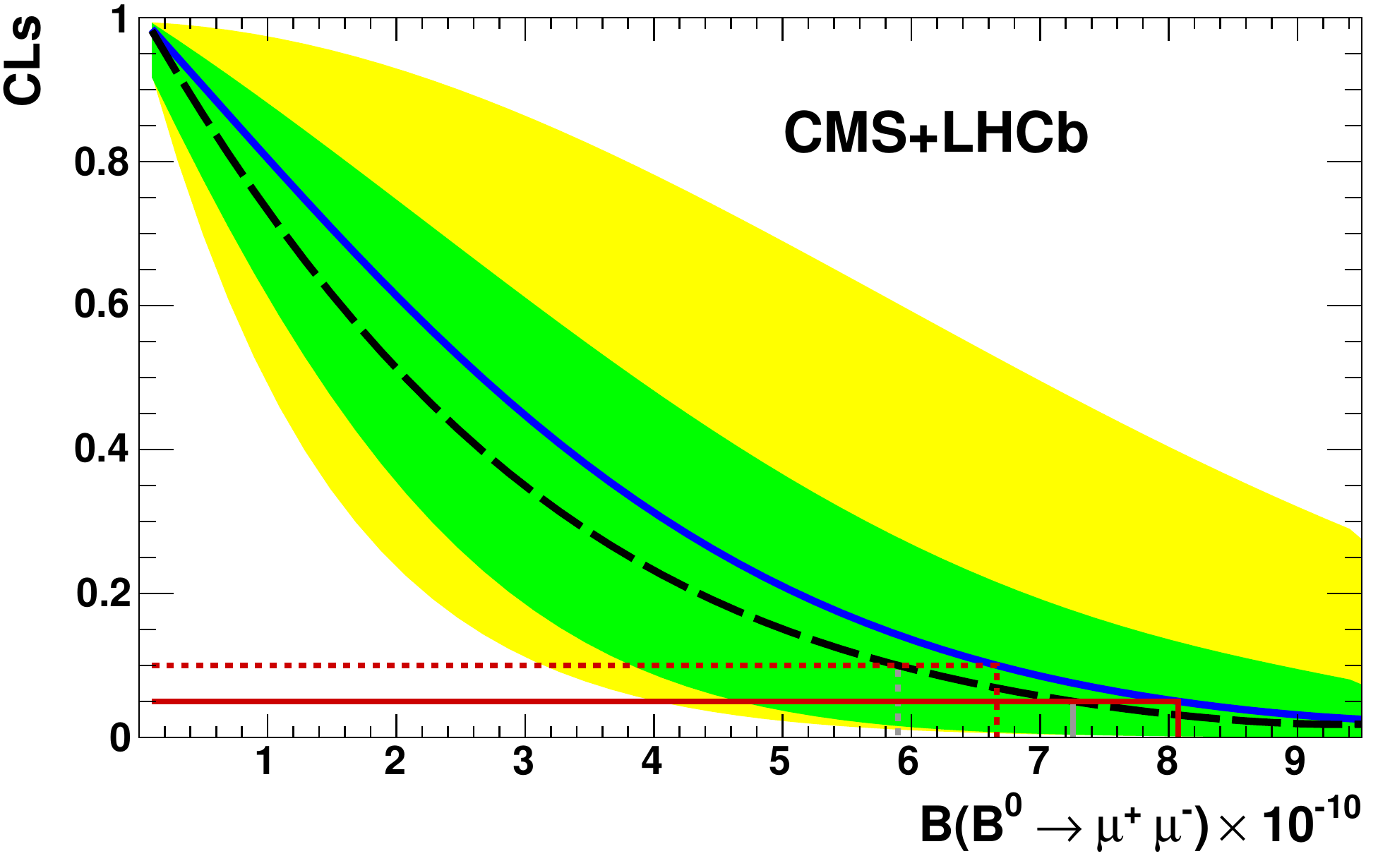}
\caption
{ \CLs as a function of the assumed \BF\ for \Bsmumu (left) and \Bdmm
  (right). 
The dashed black curves are the medians of the expected \CLs\
distributions, if background and SM signal were observed (for \Bsmm)
and in absence of signal (\Bdmm).
The green (yellow) areas cover the $\pm 1 (2) \sigma$ intervals. 
The solid blue curves are the observed \CLs. The upper limits at
90\,\% (95\,\%) \CL are indicated by the dotted (solid) horizontal
lines. Figure reproduced from Ref.~\protect\refcite{LHCb-CONF-2012-017}.}
\label{fig:CLsvsBR}
\end{figure}

The combination uses the modified frequentist
approach, as described in Sec.~\ref{sec:cls}, to combine the individual
measurements. The expected and observed \CLs values of the three
experiments are shown in Fig.~\ref{fig:CLsvsBR}~(left) as a function of the assumed
branching ratio of \Bsmm. A exclusion limit of
\BRof\Bsmm$<4.2\times10^{-9}$ is determined.  
The compatibility of the determined branching ratio with the Standard
Model value (see Sec.~\ref{sec:thexp}) is evaluated as $1-\CLsb=0.84$.
The combination of the three LHC experiments shows a moderate excess
of the branching ratio of \Bsmumu over the background only hypothesis
of $1-\CLb=0.05$.
The combined \CLs curve for the \Bdmm decay is shown in
Fig.~\ref{fig:CLsvsBR}~(right). The observed upper limit on the branching fraction is
found to be $8.1\times10^{-10}$ at 95\% CL.

 \begin{table}[h]
\tbl{Expected and observed upper limit on \BRof\Bqmumu for the
  combination of the LHC analyses, all given at 95\% CL.
  The expected limits correspond to the median cases in which only
  background or SM signal and background events were observed.} 
{\begin{tabular}{lccc}
\hline
&\multicolumn{2}{c}{Expected}&~~~~\textbf{Observed}~~~~\\
                         &~~~~bkg only~~~~  &~~~~SM + bkg~~~~ & \\
                      
\hline    
\TTstrut
~~~~\BRof{\Bsmm} $ \times 10^{-9}$~~~~& 2.3 & 6.1  & \textbf{4.2}\\ 
\BBstrut
~~~~\BRof{\Bdmm} $\times 10^{-10}$~~~~ &7.3  &      &  \textbf{8.1}\\
\hline
 \end{tabular}
\label{tab:comb}}
\end{table}

The combined upper limits on \Bsmm and \Bdmm are summarized in
Tab.~\ref{tab:comb}, they improve the limits  
obtained by the individual experiments and represent the best
existing limits on these decays. The combination improves the expected
limit with respect to the LHCb analysis alone by 15\% assuming a
signal at the SM rate any by 32\% assuming the absence of any
signal. Despite this largely improved sensitivity, the effect in the
observed upper limit is reduced as the LHCb experiment observes a
downward fluctuation of signal events with respect to the SM
prediction.

\section{Implications}
\label{sec:impl}

The exclusion limit on the branching fraction of the decay \Bsmm 
only 20\% above the SM expectation provides stringent tests of
possible extensions 
of the Standard Model. Especially the scalar sector, as discussed in
Sec.~\ref{sec:th}, is strongly constrained by the exclusion of large
enhancements in the branching fraction. 

The implications of the existing measurements are discussed in the
context of two simplified variants of the minimal supersymmetric
extension of the Standard Model (MSSM)\cite{Nilles19841,Haber198575}:
the constrained MSSM (CMSSM)\cite{Drees:1992am} and non-universal Higgs Masses of 
type 1 (NUHM1)\cite{Ellis:2008eu}. In the CMSSM, the number of free parameters is
reduced to five: $m_0, m_{1/2}$ and $A_0$, denoting common scalar, fermionic 
and trilinear soft supersymmetry-breaking parameters at the GUT scale,
and $\tan\beta$, sgn\,$\mu$ denoting the ratio of the vacuum
expectation values of the two Higgs fields and the sign of the mass
term respectively.  
In the NUHM1 model, the universality condition for the Higgs
bosons are decoupled from the other scalars, adding two extra
parameters to the model.

\begin{figure}[t]
\centering
\includegraphics[width=0.97\textwidth]{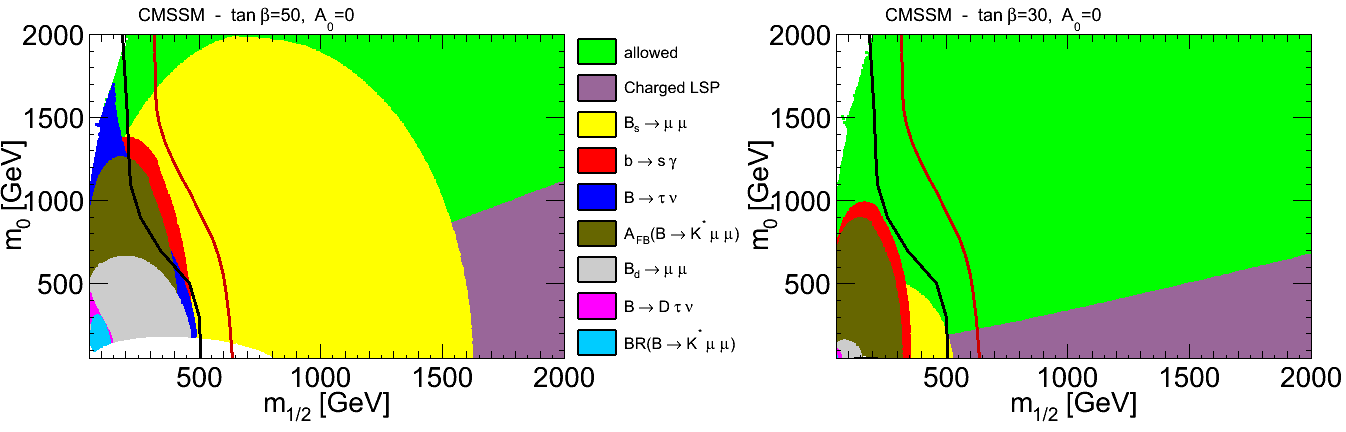}
\vspace{-2mm}
\caption{Constraints from flavor observables in CMSSM in the plane
  ($m_{1/2}, m_0$) with $A_0=0$, for $\tan\beta$ = 50 in the left and
  30 in the right. The black line corresponds to the CMS exclusion
  limit with 1.1\invfb of data and the red line to the CMS exclusion
  limit  with 4.4\invfb of data. Figures reproduced from
  Ref.~\protect\refcite{Mahmoudi:2012uk}.}  
  \label{fig:cmssm1}
\end{figure}

A analysis of the constrains imposed by the different flavor
observables in the CMSSM model is done, for example, in
Refs.~\refcite{Mahmoudi:2012un} or~\refcite{Mahmoudi:2012uk}. 
\figref{fig:cmssm1} shows the plane ($m_{1/2}, m_0$) for large (left) and
moderate (right) values of $\tan\beta$ in the CMSSM. Direct search
limits from CMS are superimposed for comparison. It can be seen that,
at large values of $\tan\beta$, the constraints from \Bsmumu are
stronger than those from direct searches. At smaller values of 
$\tan\beta$, the flavor 
observables start to lose importance compared to direct searches.
In this regime the other flavor observables, in particular the
observables measured in $\Bd \to K^* \mu^+\mu^-$ decays, loose less
sensitivity and hence play a complementary role.
In more general MSSM models, the parameter space is significantly less
constrained. 

The influence of the searches for \Bqmumu on various supersymmetric
models can be seen in a global analysis of the constraints on these
models from the measurements in the high \pt and in the 
flavor sector\cite{Buchmueller:2011sw}. A global fit to all input excluding
\BRof\Bsmm predicts in the CMSSM an enhancement up to a factor of two
and in NUHM1 up to factor of three. The current limits on \Bsmm exclude
the minimum of the global fits in both the CMSSM and NUHM1 case. 

Another aspect is the interplay between searches for \Bqmumu and Higgs
physics, since any viable model point has to be in agreement with all
direct and indirect measurements. For example, the Higgs boson
recently discovered\cite{atlas_higgs,cms_higgs} at a mass of about
$125\gevcc$ rules  
out MSSM scenarios in which the signal corresponds to the heaviest
CP-even Higgs (as opposed to the lightest Higgs), as this would imply
a too light pseudo-scalar Higgs to be consistent with the existing
constraints on \BRof\Bsmm. See Ref.~\refcite{Arbey:2012dq} for a more
detailed discussion. 

It is clear that with more precise measurements on the branching
fraction of \Bqmumu, large parts of the supersymmetric parameter space
could be disfavoured, in particular the region of high values of $\tan\beta$.
Because of the difficulty to access this range in direct searches, the
measurements of \Bqmumu are a crucial element in the exploration or
exclusion of supersymmetry. 
Also, a measurement of BR(\Bsmumu) lower
than the SM prediction would rule out a large variety of
supersymmetric models.


Once an evidence for the decay \Bsmm is found, the ratio of the
branching fraction of the \Bsmm and the \Bdmm decays becomes
accessible. This ratio is essentially free of hadronic 
uncertainties and is predicted to be\cite{Buras:2003td} 
\begin{equation}
\frac{\BRof\Bsmm}{\BRof\Bdmm}=
\frac{\hat{B}_{B_d}}{\hat{B}_{B_s}}
\frac{\tau_{B_s}}{\tau_{B_d}}
\frac{\Delta M_s}{\Delta M_d} ~~\stackrel{\rm MFV}{\approx} ~~32\, ,
\end{equation}
where the lifetime $\tau_{B_{s,d}}$ and mixing frequency $\Delta
M_{s,d}$ are experimental input and ratio of the bag parameters is
well known, $\hat{B}_{B_d}/\hat{B}_{B_s}=0.95\pm0.06$. A measurement of
this ratio provides a stringent test of the minimal flavor violation
(MFV) hypothesis\cite{D'Ambrosio:2002ex}.

\begin{figure}[h]
\centering
\includegraphics[width=0.6\textwidth]{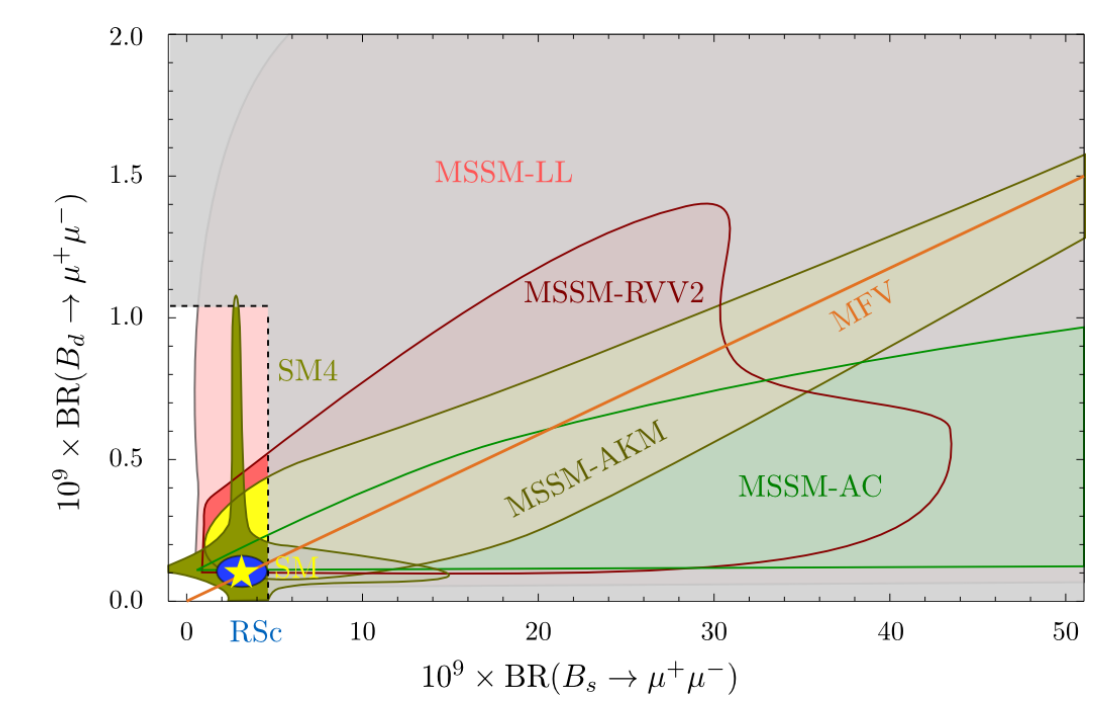}
\vspace{-2mm}
\caption{Correlation between the branching fractions of the \Bsmm and
  \Bdmm decays in MFV, the SM4 and four SUSY flavor models. The SM
  point is marked by a star and the grey area is excluded by the LHCb
  measurement given in Eq.~(\ref{eq:lhcbBs}) and~(\ref{eq:lhcbBd}).
  Figure reproduced from Ref.~\protect\refcite{Straub:2012jb}.}   
  \label{fig:mfv}
\end{figure}

The correlation between both
decays also allows to discriminate between classes of new physics
models, as shown in Fig.~\ref{fig:mfv}. A large part of the parameter
space of the supersymmetric models, where $\tan\beta$ can be large, is
ruled out by the constraints. However, in models where NP enters via
the semi-leptonic operators $O_{10}^{(\prime)}$, such as the Standard
Model with a sequential fourth generation (SM4) or Randall-Sundrum
models (RSc) are starting to be probed only now. See
Ref.~\refcite{Altmannshofer:2009ne} for a detailed discussion of all
models shown in Fig.~\ref{fig:mfv}.

\section{Summary}

The experimental situation of the searches for the very rare decays
\Bsmm and \Bdmm has been reviewed. 
The data contains a moderate excess of \Bsmm signal candidates with a
p-value of the background only hypothesis of 0.05.
The current most precise upper
exclusion limits from a single experiment are measured by the LHCb
collaboration as \BRof\Bsmm$<4.5\times 10^{-9}$, the combination of the
three LHC experiments is even more stringent at $4.2\times 10^{-9}$.
This limit is only 20\% above the SM prediction, which puts tight
constraints on various extensions of the Standard Model, especially on
supersymmetric models at high values of $\tan \beta$. 
The decay \Bdmm is constrained by the existing measurements to an
upper exclusion limit of $8\times 10^{-10}$.

The tight experimental bounds on \BRof\Bsmm precludes the optimistic
predictions by various supersymmetric models of enhancements of
several orders of magnitude. However, a possible exclusion of the
Standard Model rate would still be a clear indication of New Physics.  
It needs also to be noted that the bounds on \BRof\Bdmm
are still one order of magnitude above the predicted SM value, which
means that the most promising channel to search for New 
Physics could be the \Bd and not the \Bs decay.

The CMS and LHCb collaborations have excellent prospects to observe
the decay \Bsmm with the dataset collected in 2012. This observation,
and the precision measurement of \BRof\Bsmm in the coming years
will allow to strongly constrain the scalar sector of any extension of
the Standard Model. The next step will be to limit and later measure
the ratio of the decay rates of \Bsmm/\Bdmm, which allows a stringent
test of the hypothesis of minimal flavor violation and a good
discrimination between various extensions of the Standard Model.
\section*{Acknowledgments}


I would like to thank many of my colleagues from LHCb for pleasant and
fruitful collaboration, specially Frederic Teubert and Diego Martinez
Santos for countless discussions about rare decay analyses and Giulia
Manca for insightful discussions about production cross sections. Special thanks
also to Marco Gersabeck for proofreading this manuscript.

Furthermore, I acknowledge the support of a
Marie Curie Action: “Cofunding of the CERN Fellowship Programme
(COFUND-CERN)” of the European Community’s Seventh Framework Programme
under contract number (PCOFUND-GA-2008-229600).     

\bibliographystyle{utphys}       
\bibliography{article}   

\providecommand{\href}[2]{#2}\begingroup\raggedright\begin{thebibliography}{10}

\bibitem{Choudhury:1998ze}
S.~R. Choudhury and N.~Gaur
  \href{http://dx.doi.org/10.1016/S0370-2693(99)00203-8}{{\em Phys.Lett.}
  {\bfseries B451} (1999) 86--92},
\href{http://arxiv.org/abs/hep-ph/9810307}{{\ttfamily arXiv:hep-ph/9810307
  [hep-ph]}}.

\bibitem{Ellis:2005sc}
J.~R. Ellis, K.~A. Olive, and V.~C. Spanos
  \href{http://dx.doi.org/10.1016/j.physletb.2005.07.066}{{\em Phys.Lett.}
  {\bfseries B624} (2005) 47--59},
\href{http://arxiv.org/abs/hep-ph/0504196}{{\ttfamily arXiv:hep-ph/0504196
  [hep-ph]}}.

\bibitem{Carena:2006ai}
M.~S. Carena, A.~Menon, R.~Noriega-Papaqui, A.~Szynkman, and C.~Wagner
  \href{http://dx.doi.org/10.1103/PhysRevD.74.015009}{{\em Phys.Rev.}
  {\bfseries D74} (2006) 015009},
\href{http://arxiv.org/abs/hep-ph/0603106}{{\ttfamily arXiv:hep-ph/0603106
  [hep-ph]}}.

\bibitem{Ellis:2007ss}
J.~R. Ellis, S.~Heinemeyer, K.~Olive, and G.~Weiglein
  \href{http://dx.doi.org/10.1016/j.physletb.2007.07.056}{{\em Phys.Lett.}
  {\bfseries B653} (2007) 292--299},
\href{http://arxiv.org/abs/0706.0977}{{\ttfamily arXiv:0706.0977 [hep-ph]}}.

\bibitem{Mahmoudi:2007gd}
F.~Mahmoudi \href{http://dx.doi.org/10.1088/1126-6708/2007/12/026}{{\em JHEP}
  {\bfseries 0712} (2007) 026},
\href{http://arxiv.org/abs/0710.3791}{{\ttfamily arXiv:0710.3791 [hep-ph]}}.

\bibitem{Golowich:2011cx}
E.~Golowich, J.~Hewett, S.~Pakvasa, A.~A. Petrov, and G.~K. Yeghiyan
  \href{http://dx.doi.org/10.1103/PhysRevD.83.114017}{{\em Phys.Rev.}
  {\bfseries D83} (2011) 114017},
\href{http://arxiv.org/abs/1102.0009}{{\ttfamily arXiv:1102.0009 [hep-ph]}}.

\bibitem{Akeroyd:2011kd}
A.~Akeroyd, F.~Mahmoudi, and D.~Martinez~Santos {\em JHEP} {\bfseries 1112}
  (2011) 088,
\href{http://arxiv.org/abs/1108.3018}{{\ttfamily arXiv:1108.3018 [hep-ph]}}.

\bibitem{Huang:1998vb}
C.-S. Huang, W.~Liao, and Q.-S. Yan
  \href{http://dx.doi.org/10.1103/PhysRevD.59.011701}{{\em Phys.Rev.}
  {\bfseries D59} (1999) 011701},
\href{http://arxiv.org/abs/hep-ph/9803460}{{\ttfamily arXiv:hep-ph/9803460
  [hep-ph]}}.

\bibitem{Babu:1999hn}
K.~Babu and C.~F. Kolda
  \href{http://dx.doi.org/10.1103/PhysRevLett.84.228}{{\em Phys.Rev.Lett.}
  {\bfseries 84} (2000) 228--231},
\href{http://arxiv.org/abs/hep-ph/9909476}{{\ttfamily arXiv:hep-ph/9909476
  [hep-ph]}}.

\bibitem{Hamzaoui:1998nu}
C.~Hamzaoui, M.~Pospelov, and M.~Toharia
  \href{http://dx.doi.org/10.1103/PhysRevD.59.095005}{{\em Phys.Rev.}
  {\bfseries D59} (1999) 095005},
\href{http://arxiv.org/abs/hep-ph/9807350}{{\ttfamily arXiv:hep-ph/9807350
  [hep-ph]}}.

\bibitem{Huang:2000sm}
C.-S. Huang, W.~Liao, Q.-S. Yan, and S.-H. Zhu
  \href{http://dx.doi.org/10.1103/PhysRevD.64.059902,
  10.1103/PhysRevD.63.114021}{{\em Phys.Rev.} {\bfseries D63} (2001) 114021},
\href{http://arxiv.org/abs/hep-ph/0006250}{{\ttfamily arXiv:hep-ph/0006250
  [hep-ph]}}.

\bibitem{Blanke:2006eb}
M.~Blanke, A.~J. Buras, A.~Poschenrieder, S.~Recksiegel, C.~Tarantino, {\em
  et~al.} \href{http://dx.doi.org/10.1088/1126-6708/2007/01/066}{{\em JHEP}
  {\bfseries 0701} (2007) 066},
\href{http://arxiv.org/abs/hep-ph/0610298}{{\ttfamily arXiv:hep-ph/0610298
  [hep-ph]}}.

\bibitem{Blanke:2008yr}
M.~Blanke, A.~J. Buras, B.~Duling, K.~Gemmler, and S.~Gori
  \href{http://dx.doi.org/10.1088/1126-6708/2009/03/108}{{\em JHEP} {\bfseries
  0903} (2009) 108},
\href{http://arxiv.org/abs/0812.3803}{{\ttfamily arXiv:0812.3803 [hep-ph]}}.

\bibitem{Liu:2009hv}
W.~Liu, C.-X. Yue, and H.-D. Yang
  \href{http://dx.doi.org/10.1103/PhysRevD.79.034008}{{\em Phys.Rev.}
  {\bfseries D79} (2009) 034008},
\href{http://arxiv.org/abs/0901.3463}{{\ttfamily arXiv:0901.3463 [hep-ph]}}.

\bibitem{Bauer:2009cf}
M.~Bauer, S.~Casagrande, U.~Haisch, and M.~Neubert
  \href{http://dx.doi.org/10.1007/JHEP09(2010)017}{{\em JHEP} {\bfseries 1009}
  (2010) 017},
\href{http://arxiv.org/abs/0912.1625}{{\ttfamily arXiv:0912.1625 [hep-ph]}}.

\bibitem{Buras:2010cp}
A.~J. Buras, B.~Duling, T.~Feldmann, T.~Heidsieck, and C.~Promberger
  \href{http://dx.doi.org/10.1007/JHEP09(2010)104}{{\em JHEP} {\bfseries 1009}
  (2010) 104},
\href{http://arxiv.org/abs/1006.5356}{{\ttfamily arXiv:1006.5356 [hep-ph]}}.

\bibitem{Buras:2012ts}
A.~J. Buras and J.~Girrbach
\href{http://arxiv.org/abs/1204.5064}{{\ttfamily arXiv:1204.5064 [hep-ph]}}.

\bibitem{Chankowski:2002wr}
P.~H. Chankowski and J.~Rosiek {\em Acta Phys.Polon.} {\bfseries B33} (2002)
  2329--2354,
\href{http://arxiv.org/abs/hep-ph/0207242}{{\ttfamily arXiv:hep-ph/0207242
  [hep-ph]}}.

\bibitem{Arnowitt2002121}
R.~Arnowitt, B.~Dutta, T.~Kamon, and M.~Tanaka
  \href{http://dx.doi.org/10.1016/S0370-2693(02)01972-X}{{\em Physics Letters
  B} {\bfseries 538} no.~1–2, (2002) 121 -- 129}.

\bibitem{Fleischer:own}
R.~Fleischer {\em Int.J.Mod.Phys} {\bfseries A12} (1997) 2459.

\bibitem{Bobeth:2001sq}
C.~Bobeth, T.~Ewerth, F.~Kruger, and J.~Urban
  \href{http://dx.doi.org/10.1103/PhysRevD.64.074014}{{\em Phys.Rev.}
  {\bfseries D64} (2001) 074014},
\href{http://arxiv.org/abs/hep-ph/0104284}{{\ttfamily arXiv:hep-ph/0104284
  [hep-ph]}}.

\bibitem{Altmannshofer:2011gn}
W.~Altmannshofer, P.~Paradisi, and D.~M. Straub
  \href{http://dx.doi.org/10.1007/JHEP04(2012)008}{{\em JHEP} {\bfseries 1204}
  (2012) 008},
\href{http://arxiv.org/abs/1111.1257}{{\ttfamily arXiv:1111.1257 [hep-ph]}}.

\bibitem{Mahmoudi:2012un}
F.~Mahmoudi, S.~Neshatpour, and J.~Orloff
\href{http://arxiv.org/abs/1205.1845}{{\ttfamily arXiv:1205.1845 [hep-ph]}}.

\bibitem{Laiho:2009eu}
J.~Laiho, E.~Lunghi, and R.~S. Van~de Water
  \href{http://dx.doi.org/10.1103/PhysRevD.81.034503}{{\em Phys.Rev.}
  {\bfseries D81} (2010) 034503},
\href{http://arxiv.org/abs/0910.2928}{{\ttfamily arXiv:0910.2928 [hep-ph]}}.

\bibitem{Gamiz:2009ku}
{\bfseries HPQCD collaboration}, E.~Gamiz, C.~T. Davies, G.~P. Lepage,
  J.~Shigemitsu, and M.~Wingate
  \href{http://dx.doi.org/10.1103/PhysRevD.80.014503}{{\em Phys.Rev.}
  {\bfseries D80} (2009) 014503},
\href{http://arxiv.org/abs/0902.1815}{{\ttfamily arXiv:0902.1815 [hep-lat]}}.

\bibitem{Na:2012kp}
H.~Na, C.~J. Monahan, C.~T. Davies, R.~Horgan, G.~P. Lepage, {\em et~al.}
\href{http://arxiv.org/abs/1202.4914}{{\ttfamily arXiv:1202.4914 [hep-lat]}}.

\bibitem{Davies:2012qf}
C.~Davies {\em PoS} {\bfseries LATTICE2011} (2011) 019,
\href{http://arxiv.org/abs/1203.3862}{{\ttfamily arXiv:1203.3862 [hep-lat]}}.

\bibitem{Charles:2011va}
J.~Charles, O.~Deschamps, S.~Descotes-Genon, R.~Itoh, H.~Lacker, {\em et~al.}
  \href{http://dx.doi.org/10.1103/PhysRevD.84.033005}{{\em Phys.Rev.}
  {\bfseries D84} (2011) 033005},
\href{http://arxiv.org/abs/1106.4041}{{\ttfamily arXiv:1106.4041 [hep-ph]}}.

\bibitem{utfit}
{\bfseries UTFit collaboration} {\em http://utfit.org} .

\bibitem{Buras:2003td}
A.~J. Buras \href{http://dx.doi.org/10.1016/S0370-2693(03)00561-6}{{\em
  Phys.Lett.} {\bfseries B566} (2003) 115--119},
\href{http://arxiv.org/abs/hep-ph/0303060}{{\ttfamily arXiv:hep-ph/0303060
  [hep-ph]}}.

\bibitem{Buras:2003jf}
A.~J. Buras {\em Acta Phys.Polon.} {\bfseries B34} (2003) 5615--5668,
\href{http://arxiv.org/abs/hep-ph/0310208}{{\ttfamily arXiv:hep-ph/0310208
  [hep-ph]}}.

\bibitem{Shigemitsu:2009jy}
{\bfseries HPQCD collaboration}, J.~Shigemitsu {\em et~al.} {\em PoS}
  {\bfseries LAT2009} (2009) 251,
\href{http://arxiv.org/abs/0910.4131}{{\ttfamily arXiv:0910.4131 [hep-lat]}}.

\bibitem{LHCb-CONF-2012-002}
{\bfseries LHCb collaboration}. LHCb-CONF-2012-002.

\bibitem{deBruyn:2012wk}
K.~de~Bruyn, R.~Fleischer, R.~Knegjens, P.~Koppenburg, M.~Merk, {\em et~al.}
\href{http://arxiv.org/abs/1204.1737}{{\ttfamily arXiv:1204.1737 [hep-ph]}}.

\bibitem{deBruyn:2012wj}
K.~de~Bruyn, R.~Fleischer, R.~Knegjens, P.~Koppenburg, M.~Merk, {\em et~al.}
\href{http://arxiv.org/abs/1204.1735}{{\ttfamily arXiv:1204.1735 [hep-ph]}}.

\bibitem{babarPhys}
{\bfseries BABAR collaboration}. SLAC Report SLAC-R-504.

\bibitem{PhysRevD.75.012010}
{\bfseries CDF collaboration}
  \href{http://dx.doi.org/10.1103/PhysRevD.75.012010}{{\em Phys. Rev. D}
  {\bfseries 75} (Jan, 2007) 012010}.

\bibitem{Aaij:2012jd}
{\bfseries LHCb collaboration}, R.~Aaij {\em et~al.}
  \href{http://dx.doi.org/10.1007/JHEP04(2012)093}{{\em JHEP} {\bfseries 1204}
  (2012) 093},
\href{http://arxiv.org/abs/1202.4812}{{\ttfamily arXiv:1202.4812 [hep-ex]}}.

\bibitem{Chatrchyan:2011pw}
{\bfseries CMS collaboration}, S.~Chatrchyan {\em et~al.}
  \href{http://dx.doi.org/10.1103/PhysRevLett.106.252001}{{\em Phys.Rev.Lett.}
  {\bfseries 106} (2011) 252001},
\href{http://arxiv.org/abs/1104.2892}{{\ttfamily arXiv:1104.2892 [hep-ex]}}.

\bibitem{PhysRevD.85.032008}
{\bfseries LHCb collaboration}, e.~a. Aaji
  \href{http://dx.doi.org/10.1103/PhysRevD.85.032008}{{\em Phys. Rev. D}
  {\bfseries 85} (Feb, 2012) 032008}.

\bibitem{Aaltonen:2011fi}
{\bfseries CDF collaboration}, T.~Aaltonen {\em et~al.}
  \href{http://dx.doi.org/10.1103/PhysRevLett.107.191801,
  10.1103/PhysRevLett.107.239903, 10.1103/PhysRevLett.107.191801,
  10.1103/PhysRevLett.107.239903}{{\em Phys.Rev.Lett.} {\bfseries 107} (2011)
  239903}, \href{http://arxiv.org/abs/1107.2304}{{\ttfamily arXiv:1107.2304
  [hep-ex]}}.
7 pages, 1 figure/ version accepted by PRL.

\bibitem{cdf_10fb}
{\bfseries CDF collaboration} {\em
  http://www-cdf.fnal.gov/physics/new/bottom/120209.blessed-bmumu10fb/} .

\bibitem{Abazov:2010fs}
{\bfseries D0 collaboration}, V.~M. Abazov {\em et~al.}
  \href{http://dx.doi.org/10.1016/j.physletb.2010.09.024}{{\em Phys.Lett.}
  {\bfseries B693} (2010) 539--544},
\href{http://arxiv.org/abs/1006.3469}{{\ttfamily arXiv:1006.3469 [hep-ex]}}.

\bibitem{Aaij:2011rja}
{\bfseries LHCb collaboration}, R.~Aaij {\em et~al.}
  \href{http://dx.doi.org/10.1016/j.physletb.2011.04.031}{{\em Phys.Lett.}
  {\bfseries B699} (2011) 330--340},
\href{http://arxiv.org/abs/1103.2465}{{\ttfamily arXiv:1103.2465 [hep-ex]}}.

\bibitem{LHCb:2011ac}
{\bfseries LHCb collaboration}, R.~Aaij {\em et~al.}
  \href{http://dx.doi.org/10.1016/j.physletb.2012.01.038}{{\em Phys.Lett.}
  {\bfseries B708} (2012) 55--67},
\href{http://arxiv.org/abs/1112.1600}{{\ttfamily arXiv:1112.1600 [hep-ex]}}.

\bibitem{Aaij:2012ac}
{\bfseries LHCb collaboration}, R.~Aaij {\em et~al.}
  \href{http://dx.doi.org/10.1103/PhysRevLett.108.231801}{{\em Phys. Rev. Lett.
  108,} {\bfseries 231801} (2012) },
\href{http://arxiv.org/abs/1203.4493}{{\ttfamily arXiv:1203.4493 [hep-ex]}}.

\bibitem{Aad:2012pn}
{\bfseries ATLAS collaboration}, G.~Aad {\em et~al.}
  \href{http://dx.doi.org/10.1016/j.physletb.2012.06.013}{{\em Phys. Lett.}
  {\bfseries B713} (2012) 387},
\href{http://arxiv.org/abs/1204.0735}{{\ttfamily arXiv:1204.0735 [hep-ex]}}.

\bibitem{Chatrchyan:2011kr}
{\bfseries CMS collaboration}, S.~Chatrchyan {\em et~al.}
  \href{http://dx.doi.org/10.1103/PhysRevLett.107.191802}{{\em Phys.Rev.Lett.}
  {\bfseries 107} (2011) 191802},
\href{http://arxiv.org/abs/1107.5834}{{\ttfamily arXiv:1107.5834 [hep-ex]}}.

\bibitem{Chatrchyan:2012rg}
{\bfseries CMS collaboration}, S.~Chatrchyan {\em et~al.}
\href{http://arxiv.org/abs/1203.3976}{{\ttfamily arXiv:1203.3976 [hep-ex]}}.

\bibitem{Aad:2012xs}
{\bfseries ATLAS collaboration}, G.~Aad {\em et~al.}
  \href{http://dx.doi.org/10.1140/epjc/s10052-011-1849-1}{{\em Eur.Phys.J.}
  {\bfseries C72} (2012) 1849},
\href{http://arxiv.org/abs/1110.1530}{{\ttfamily arXiv:1110.1530 [hep-ex]}}.

\bibitem{Aaij:1384386}
R.~Aaij and J.~Albrecht Tech. Rep. LHCb-PUB-2011-017. CERN-LHCb-PUB-2011-017,
  CERN, Geneva, Sep, 2011.

\bibitem{pdg10}
{\bfseries Particle Data Group}, {K. Nakamura et al.} {\em J. Phys. G}
  {\bfseries 37} (2010) 075021.

\bibitem{Louvot:2009xg}
{\bfseries Belle collaboration}, R.~Louvot
\href{http://arxiv.org/abs/0905.4345}{{\ttfamily arXiv:0905.4345 [hep-ex]}}.

\bibitem{Junk:1999kv}
T.~Junk \href{http://dx.doi.org/10.1016/S0168-9002(99)00498-2}{{\em
  Nucl.Instrum.Meth.} {\bfseries A434} (1999) 435--443},
\href{http://arxiv.org/abs/hep-ex/9902006}{{\ttfamily arXiv:hep-ex/9902006
  [hep-ex]}}.

\bibitem{0954-3899-28-10-313}
A.~L. Read {\em Journal of Physics G: Nuclear and Particle Physics} {\bfseries
  28} no.~10, (2002) 2693.

\bibitem{LHCb-CONF-2012-017}
{\bfseries ATLAS, CMS and LHCb collaborations} {\em LHCb-CONF-2012-017} (May,
  2012) .

\bibitem{Nilles19841}
H.~Nilles \href{http://dx.doi.org/10.1016/0370-1573(84)90008-5}{{\em Physics
  Reports} {\bfseries 110} no.~1–2, (1984) 1 -- 162}.

\bibitem{Haber198575}
H.~Haber and G.~Kane \href{http://dx.doi.org/10.1016/0370-1573(85)90051-1}{{\em
  Physics Reports} {\bfseries 117} no.~2–4, (1985) 75 -- 263}.

\bibitem{Drees:1992am}
M.~Drees and M.~M. Nojiri \href{http://dx.doi.org/10.1103/PhysRevD.47.376}{{\em
  Phys.Rev.} {\bfseries D47} (1993) 376--408},
\href{http://arxiv.org/abs/hep-ph/9207234}{{\ttfamily arXiv:hep-ph/9207234
  [hep-ph]}}.

\bibitem{Ellis:2008eu}
J.~R. Ellis, K.~A. Olive, and P.~Sandick
  \href{http://dx.doi.org/10.1103/PhysRevD.78.075012}{{\em Phys.Rev.}
  {\bfseries D78} (2008) 075012},
\href{http://arxiv.org/abs/0805.2343}{{\ttfamily arXiv:0805.2343 [hep-ph]}}.

\bibitem{Mahmoudi:2012uk}
F.~Mahmoudi \href{http://arxiv.org/abs/1205.3099}{{\ttfamily arXiv:1205.3099
  [hep-ph]}}.
Proceedings of Moriond QCD 2012.

\bibitem{Buchmueller:2011sw}
O.~Buchmueller, R.~Cavanaugh, A.~De~Roeck, M.~Dolan, J.~Ellis, {\em et~al.}
  \href{http://dx.doi.org/10.1140/epjc/s10052-012-1878-4}{{\em Eur.Phys.J.}
  {\bfseries C72} (2012) 1878},
\href{http://arxiv.org/abs/1110.3568}{{\ttfamily arXiv:1110.3568 [hep-ph]}}.

\bibitem{atlas_higgs}
{\bfseries ATLAS collaboration}, F.~Gianotti {\em CERN Seminar, July 4th, 2012}
  .

\bibitem{cms_higgs}
{\bfseries CMS collaboration}, J.~Incandela {\em CERN Seminar, July 4th, 2012}
  .

\bibitem{Arbey:2012dq}
A.~Arbey, M.~Battaglia, A.~Djouadi, and F.~Mahmoudi
\href{http://arxiv.org/abs/1207.1348}{{\ttfamily arXiv:1207.1348 [hep-ph]}}.

\bibitem{D'Ambrosio:2002ex}
G.~D'Ambrosio, G.~Giudice, G.~Isidori, and A.~Strumia
  \href{http://dx.doi.org/10.1016/S0550-3213(02)00836-2}{{\em Nucl.Phys.}
  {\bfseries B645} (2002) 155--187},
\href{http://arxiv.org/abs/hep-ph/0207036}{{\ttfamily arXiv:hep-ph/0207036
  [hep-ph]}}.

\bibitem{Altmannshofer:2009ne}
W.~Altmannshofer, A.~J. Buras, S.~Gori, P.~Paradisi, and D.~M. Straub
  \href{http://dx.doi.org/10.1016/j.nuclphysb.2009.12.019}{{\em Nucl.Phys.}
  {\bfseries B830} (2010) 17--94},
\href{http://arxiv.org/abs/0909.1333}{{\ttfamily arXiv:0909.1333 [hep-ph]}}.

\bibitem{Straub:2012jb}
D.~M. Straub
\href{http://arxiv.org/abs/1205.6094}{{\ttfamily arXiv:1205.6094 [hep-ph]}}.

\end{thebibliography}\endgroup

\end{document}